\documentclass[12pt]{article}
\usepackage{amssymb}
\usepackage{amsmath,bm}
\usepackage{graphics,mathrsfs}
\usepackage{graphicx,epsfig}
\usepackage{subfigure}
\usepackage{setspace}
\usepackage{enumerate}
\usepackage{enumitem}
\usepackage{caption}
\usepackage{varioref}
\usepackage{color}
\usepackage{natbib}
\usepackage{epstopdf}
\setcitestyle{authoryear,open={(},close={)}}
\usepackage{amsthm}
\makeatletter \oddsidemargin  -.1in \evensidemargin -.1in
\begin{document}
	
	\newcommand{\bea}{\begin{eqnarray}}
		\newcommand{\eea}{\end{eqnarray}}
	\newcommand{\nn}{\nonumber}
	\newcommand{\bee}{\begin{eqnarray*}}
		\newcommand{\eee}{\end{eqnarray*}}
	\newcommand{\lb}{\label}
	\newcommand{\nii}{\noindent}
	\newcommand{\ii}{\indent}
	\newtheorem{theorem}{Theorem}[section]
	\newtheorem{example}{Example}[section]
	\newtheorem{counterexample}{Counterexample}[section]
	\newtheorem{corollary}{Corollary}[section]
	\newtheorem{definition}{Definition}[section]
	\newtheorem{lemma}{Lemma}[section]
	\newtheorem{remark}{Remark}[section]
	\newtheorem{notation}{Notation}[section]
	\newtheorem{proposition}{Proposition}[section]
	\renewcommand{\theequation}{\thesection.\arabic{equation}}
	\renewcommand{\labelenumi}{(\roman{enumi})}
	\title{\bf Estimation of parameters of the logistic exponential distribution under progressive type-I hybrid censored sample*}
	\author{ Subhankar {\bf Dutta}\thanks { 
			Email address: subhankar.dta@gmail.com} ~and  Suchandan {\bf  Kayal}\thanks {Email address (corresponding author):
			kayals@nitrkl.ac.in,~suchandan.kayal@gmail.com, It has been published in Quality Technology \& Quantitative Management, https://doi.org/10.1080/16843703.2022.2027601}
		\\{\it \small Department of Mathematics, National Institute of
			Technology Rourkela, Rourkela-769008, India}}
	\date{}
	\maketitle
	\begin{center}
		{\large \bf Abstract}
	\end{center}
	The present paper addresses the problem of estimation of the model
	parameters of the logistic exponential distribution based on progressive type-I
	hybrid censored sample. The maximum likelihood estimates
	are obtained and computed numerically using Newton-Raphson algorithm.
	Further, the Bayes estimates are derived under the squared error, LINEX
	and the generalized entropy loss functions. Two types (independent and
	bivariate) of prior distributions are considered for the purpose of
	Bayesian estimation. It is seen that the Bayes estimates are not of explicit forms.
	Thus, Lindley's approximation technique is employed to get
	approximate Bayes estimates. Interval estimates of the parameters
	based on the normal approximation of the maximum likelihood estimates and the log-transformed maximum likelihood
	estimates are constructed. The highest posterior density credible
	intervals are obtained by using the importance sampling method. Numerical simulation is
	performed to see the performance of the proposed estimation techniques. A real life data set is considered and analysed for the purpose of
	illustrations.
	
	\section{Introduction}
	Censoring was introduced in practice to save time and to reduce the
	number of failed units associated with a life-testing experiment.
	The concept of censoring is very common in reliability and survival
	studies. The type-I and type-II censoring schemes are two
	fundamental censoring schemes among all the schemes, developed so
	far. Suppose that $n$ units are placed to conduct a life-testing
	experiment. In type-I censoring scheme, it is assumed that the
	experiment runs up to a prefixed time, say $T$. Here, the random
	number of failures, say $m~(<n)$ in the time interval $[0,T]$ is
	observed. For the case of type-II censoring scheme, the
	experiment continues till a predefined number of failures, say
	$m$. Thus, clearly the time, say $T$ at which the experiment stops
	is random. \citet{epstein1954truncated} introduced a censoring
	scheme by mixing type-I and type-II schemes, which  is
	dubbed as the hybrid censoring scheme. The main
	drawback of the hybrid censoring scheme is that it does not have
	flexibility of removing experimental units before the experiment
	stops. However, the progressive type-II censoring scheme has
	such flexibility. Due to this virtue, the progressive type-II
	censoring design has been widely utilized to analyse lifetime data.
	But, these days, the system-components are highly
	reliable. This leads to a large experimental duration in
	progressive type-II censoring scheme. To overcome such difficulty,
	\cite{kundu2006analysis} introduced a general censoring scheme,
	known as the progressive type-I hybrid censoring scheme (PT-IHCS). Note that
	the PT-IHCS is a mixture of the
	hybrid and the progressive type-II censoring schemes, which is described in the
	following.

	Suppose an experiment has been conducted with $n$ components. The random lifetimes of $n$ components are denoted by
	$X_{1},\ldots,X_{n}.$  Let an integer $m~(<n)$ and the time point
	$T$ be fixed when the experiment starts.  At the time of first
	failure, say $X_{1:m:n}$, $R_{1}$ number of live units are randomly
	taken out. When the second failure occurs at $X_{2:m:n}$, $R_{2}$
	units are removed from the experiment. Similarly, for the third
	failure, fourth failure and so on. Thus, two cases arise as follows. Assume that the $m$th
	failure occurs at time $X_{m:m:n}$, which is less than $T$. Then,
	the experiment stops at $X_{m:m:n}$. Other ways, let the $m$th
	failure occur after the time point $T$. The $j~(0\le j <m)$ number
	of failures occur before $T$. Then, at  $T$, the remaining units
	$R^{*}_{j}=n-R_{1}-R_{2}-.....-R_{j}-j$ are removed and the
	experiment terminates. These two cases are named as Case-$A$ and
	Case-$B$, which are presented below. This censoring scheme is known as the PT-IHCS. For details on this censoring design, one
	may refer to \cite{kundu2006analysis}. Note that for the progressive type-I hybrid censored sample, one has either Case-$A$ or
	Case-$B$, where
	\begin{eqnarray*}
		\mbox{Case}\mbox{-A}&:&   \{X_{1:m:n},\ldots,X_{m:m:n}\}, ~~      \mbox{if}~~ X_{m:m:n} < T;\\
		\mbox{Case}\mbox{-B}&:& \{X_{1:m:n},\ldots,X_{j:m:n}\},  ~~~      \mbox{if}~~ X_{j:m:n}<T<X_{j+1:m:n}.
	\end{eqnarray*}
	The meaning of $X_{j:m:n}<T<X_{j+1:m:n}$ is that the $j$th failure occurs before $T$, and no failure occurs between the points $X_{j:m:m}$ and $T$. That is, $X_{j+1:m:n},\ldots,X_{m:m:n}$ are not observed. Due to the importance of  PT-IHCS, many authors have studied estimation of the parameters of various lifetime distributions based on this scheme.  \cite{Lin} obtained
	some useful classical estimates of the unknown parameters of Weibull distribution
	under the PT-IHCS. The authors obtained maximum likelihood estimates (MLEs)
	and approximate MLEs.  \cite{hemmati2013statistical} obtained various estimates for the log-normal distribution under progressive type-I hybrid censored (PT-IHC) sample. \cite{sultana2019parameter} studied estimation of parameters of the generalized half-normal distribution based on the PT-IHCS. They obtained maximum likelihood and Bayes estimates. Further, the authors computed Bayes estimates using different approximation techniques. \cite{gamchi2019classical} considered classical and Bayesian inference of the Burr type-III distribution based on the PT-IHC data.  One may refer to \cite{kayal2019statistical}, \cite{goyal2020bayesian}, \cite{sultana2020inference} and among others for some recent developments in this topic.
	
	\begin{figure}[!htbp]\label{fig1.1}
				\begin{center}
					\includegraphics[width=4in]{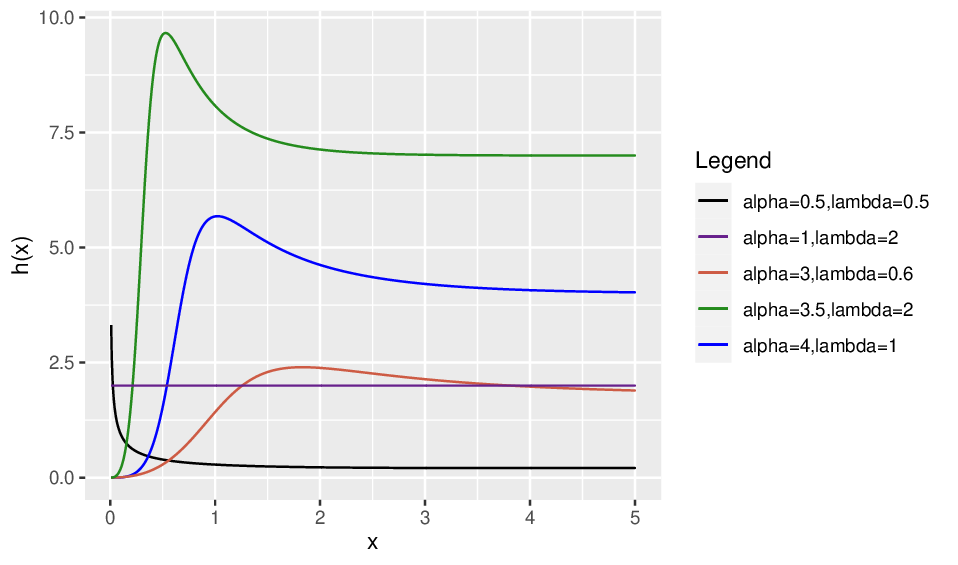}
					\caption{ Plot of the hazard rate function of LE distribution with various combinations of parameters. }
				\end{center}
			\end{figure}
	
	Let $X$ be a random variable following logistic exponential (LE)
	distribution (see \cite{Lan}) with probability density function (PDF) $f$
	and cumulative distribution function (CDF) $F$, where
	\begin{eqnarray}\label{eq1.1}
		f(x)\equiv f(x;\alpha,\lambda)= \frac{\alpha\lambda e^{\lambda x} (e^{\lambda
				x}-1)^{\alpha-1}}{\left[1+\left(e^{\lambda
				x}-1\right)^{\alpha}\right]^2};~x>0,~\alpha,~\lambda>0
	\end{eqnarray}
	and
	\begin{eqnarray}\label{eq1.2}
		F(x)\equiv F(x;\alpha,\lambda)=1-\left[1+(e^{\lambda
			x}-1)^{\alpha}\right]^{-1};~x>0,~\alpha,~\lambda>0.
	\end{eqnarray}
	Note that the LE distribution is a generalization of the
	exponential distribution. The CDF in (\ref{eq1.2}) reduces to the CDF of exponential distribution when $\alpha=1$.
	The hazard rate function of the LE distribution  is given by
	\begin{eqnarray}
		h(x)\equiv h(x;\alpha,\lambda)= \frac{\alpha \lambda  e^{\lambda x} (e^{\lambda x}-1)^{\alpha-1}}{1+\left(e^{\lambda
				x}-1\right)^{\alpha}};~x>0,~\alpha,~\lambda>0.
	\end{eqnarray}
It becomes constant when $\alpha=1$, that is, it belongs to both increasing hazard rate and decreasing hazard rate classes. The LE distribution with CDF (\ref{eq1.2}) has increasing and decreasing shaped hazard rate when $\alpha,~\lambda>1$. Further, the shape of the hazard rate function of this distribution is bathtub (upside down bathtub) when  $0<\alpha<1$ ($\alpha >1$). Please refer to \cite{Lan} for various other properties of this distribution. It is worth mentioning that the LE distribution is the only two-parameter family of distributions that has five types of shapes of the hazard rate function.  Figure $1$ depicts the graph of the hazard rate function of the LE distribution for various combinations of the parameters.

We recall that in reliability engineering, biology and several statistical modelings, different-shaped hazard rate functions are used with different interpretations. Consider a high failure rate during infancy, which decreases to a certain level, then it remains constant up to a certain time, and then it increases due to aging. This type of situation can be described by a bathtub-shaped hazard rate model. The data set on hydrodynamical devices  (see \cite{hjorth1980reliability}) can be modeled using a statistical distribution having  bathtub-shaped hazard rate. The decreasing hazard rate is usually observed in early stage of a working system. Data set on the earthquakes in North Anatolia fault zone can be fitted by a decreasing hazard model  (see \cite{kucs2007new}). Further, increasing hazard rate models are used in operations and supply chain management research due to implications in evaluation of some type of objective functions to model stochastic events. In biology, it is observed in the course of a disease whose mortality reaches a peak after some finite period and then declines gradually. Thus, the data set related to this can be modeled using upside down bathtub-shaped hazard rate model. It is already mentioned before that the LE distribution exhibits the constant, increasing, decreasing, bathtub and upside down bathtub-shaped hazard rate. Due to various shapes of the hazard rate function, the LE distribution is very flexible to use in different areas of research such as clinical, reliability and survival studies. Due to wide practical utility of the LE distribution, we consider the problem of estimation of the model parameters problem  with the importance of the PT-IHCS motivate us to consider the present estimation problem, since this problem has not been reported earlier in the literature.

First, we consider the MLEs of the unknown parameters based on the PT-IHC data from the LE distribution. The MLEs can not be obtained explicitly.  Thus, Newton-Raphson iterative method is used to solve non-linear equations. Although, the performance of the MLEs is satisfactory, another way to estimate the parameters is Bayesian estimation. For Bayesian estimation, we need to choose prior distributions on the unknown model parameters. In this paper, the independent as well as bivariate priors have been considered. As expected, the explicit expressions of the Bayes estimates can not be obtained in explicit forms. Thus, we propose to use Lindley's approximation method to compute the Bayes estimates. The importance sampling method is also used to compute the Bayes estimates and the highest posterior density (HPD) credible intervals. However, the values of the Bayes estimates using importance sampling method is not presented in the tables to maintain brevity. The approximate confidence intervals are also obtained using the asymptotic distribution of the MLEs. We consider a real life data set and analyzed it for the purpose of illustrations.

	The paper is arranged as follows. In Section $2$, the MLEs of the parameters are computed using Newton-Raphson  iterative method. To get rough idea on the existence and uniqueness of the MLEs, the plots of the profile log-likelihood function of the parameters are presented. The Bayes estimates are obtained using three loss functions. Two different types of the prior distributions are considered in this purpose. The form of the Bayes estimates are not of closed forms. Thus, Lindley's approximation method is employed in Section $4$. Further, importance sampling method is utilized in order to construct HPD credible interval estimates. In Section $5$, two methods are used to compute asymptotic confidence intervals of the parameters. A simulation study is carried out in Section $6$ to observe the comparative performance of the proposed estimates. Section $7$ deals with a real life data set. Finally, in Section $8$, some concluding remarks are added.
	
	\section{Maximum likelihood estimation\setcounter{equation}{0}}
	This section focuses on the maximum likelihood estimation of the unknown model parameters $\alpha$ and $\lambda$ of the LE distribution with CDF (\ref{eq1.2}) based on PT-IHC sample. Given the observed data, the likelihood function of $\alpha$ and $\lambda$ can be written as follows:
	\begin{eqnarray} \label{eq2.1}
		\mbox{Case-A :}~~~~~L(\alpha,\lambda) &\propto& \prod_{i=1}^{m} f(x_{i:m:n})
		[1-F(x_{i:m:n})]^{R_{i}},\\\label{eq2.2}
		\mbox{Case-B :}~~~~~L(\alpha,\lambda) &\propto& \prod_{i=1}^{j} f(x_{i:m:n})
		[1-F(x_{i:m:n})]^{R_{i}}[1-F(T)]^{R^{*}_{j}},
	\end{eqnarray}
	where $m ~(\leq n) $  is a prefixed integer and
	$R^{*}_{j}=n-R_{1}-R_{2}-\ldots-R_{j}-j$. Henceforth, for notational
	convenience, $x_{i}$ is used for $x_{i:m:n}$. Eqs.
	$(\ref{eq2.1})$ and $(\ref{eq2.2})$ can be combined as
	\begin{eqnarray}\label{eq2.3}
		L(\alpha,\lambda) \propto \prod_{i=1}^{D} f(x_{i})
		[1-F(x_{i})]^{R_{i}}[1-F(T)]^{R^{*}_{D}},
	\end{eqnarray}
	where for  Case-$A$, $D=m$, $R^{*}_{D}=0$, and for Case-$B$, $D=j$,
	$R^{*}_{D}=n-R_{1}-R_{2}-\ldots-R_{D}-D$. Denote $g(x;\lambda)=e^{\lambda x}-1,~x>0,~\lambda>0.$ Now, using (\ref{eq1.1})
	and (\ref{eq1.2}), the likelihood function in (\ref{eq2.3}) becomes
	\begin{eqnarray}\label{eq2.4}
		L(\alpha,\lambda)\propto\alpha^{D}\lambda^{D}e^{\lambda
			\sum_{i=1}^{D}x_{i}} \prod_{i=1}^{D} (g(x_i;\lambda))^{\alpha-1} [1+(g(x_i;\lambda))^{\alpha}]^{-(R_{i}+2)}
		[1+(g(T;\lambda))^{\alpha}]^{-R^{*}_{D}}.
	\end{eqnarray}
	Thus, the log-likelihood function is
	\begin{eqnarray}\label{eq2.5}
		\nonumber \log L(\alpha,\lambda) &=& D\log \alpha + D\log \lambda +
		\lambda \sum_{i=1}^{D} x_{i} + (\alpha -1)\sum_{i=1}^{D}\log
		(g(x_i;\lambda))\\ &~&- \sum_{i=1}^{D}(R_{i}+2) \log
		[1+(g(x_i;\lambda))^{\alpha}] - R^{*}_{D} \log [1+(g(T;\lambda))^{\alpha}].
	\end{eqnarray}
	Now, differentiating (\ref{eq2.5}) with respect to $\alpha$ and
	$\lambda$, and equating to zero, the likelihood equations are respectively
	obtained as
	\begin{eqnarray}
		\nonumber \frac{\partial\log L}{\partial \alpha}&=&\frac{D}{\alpha} + \sum_{i=1}^{D} \log (g(x_i;\lambda)) -\sum_{i=1}^{D}\frac{(R_{i}+2)(g(x_i;\lambda))^{\alpha}\log (g(x_i;\lambda))}{[1+(g(x_i;\lambda))^{\alpha}]}\\
		&~&-\frac{R^{*}_{D}(g(T;\lambda))^{\alpha}\log (g(T;\lambda))}{[1+(g(T;\lambda))^{\alpha}]}=0  \label{eq2.6}
	\end{eqnarray}
	and
	\begin{eqnarray}
		\nonumber \frac{\partial\log L}{\partial\lambda}&=&\frac{D}{\lambda}+ \sum_{i=1}^{D}x_{i} + (\alpha-1)\sum_{i=1}^{D}\frac{x_{i} e^{\lambda x_{i}}}{g(x_i;\lambda)}-\sum_{i=1}^{D}\frac{(R_{i}+2)\alpha x_{i}(g(x_i;\lambda))^{\alpha-1}e^{\lambda x_{i}}}{[1+(g(x_i;\lambda))^{\alpha}]}\\
		&~&-\frac{R^{*}_{D}\alpha T(g(T;\lambda))^{\alpha-1}e^{\lambda
				T}}{[1+(g(T;\lambda))^{\alpha}]}=0. \label{eq2.7}
	\end{eqnarray}
	The MLEs of $\alpha$ and $\lambda$ are the
	solution of the simultaneous  nonlinear equations given by (\ref{eq2.6}) and
	(\ref{eq2.7}). It is not easy to get solution of these equations in
	explicit form. So, Newton-Raphson (NR) iterative method is utilized to compute the
	MLEs of $\alpha$ and $\lambda$, which are
	respectively denoted by $\widehat{\alpha}$ and $\widehat{\lambda}$. Note that in order to apply NR method, "nleqslv" package in R software has been used.
	\begin{figure}[h!]
		\begin{center}
			\subfigure[]{\label{c1}\includegraphics[height=1.85in]{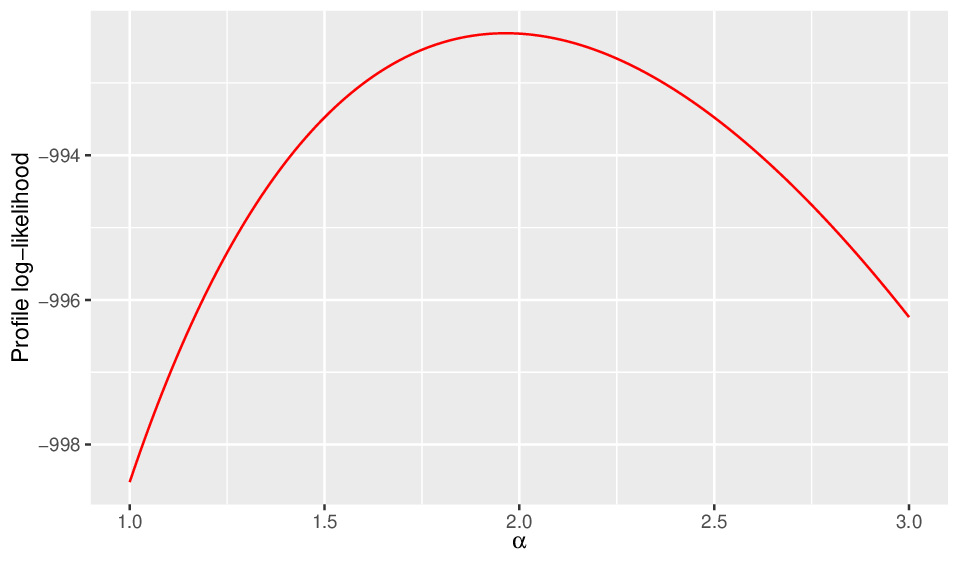}}
			\subfigure[]{\label{c1}\includegraphics[height=1.85in]{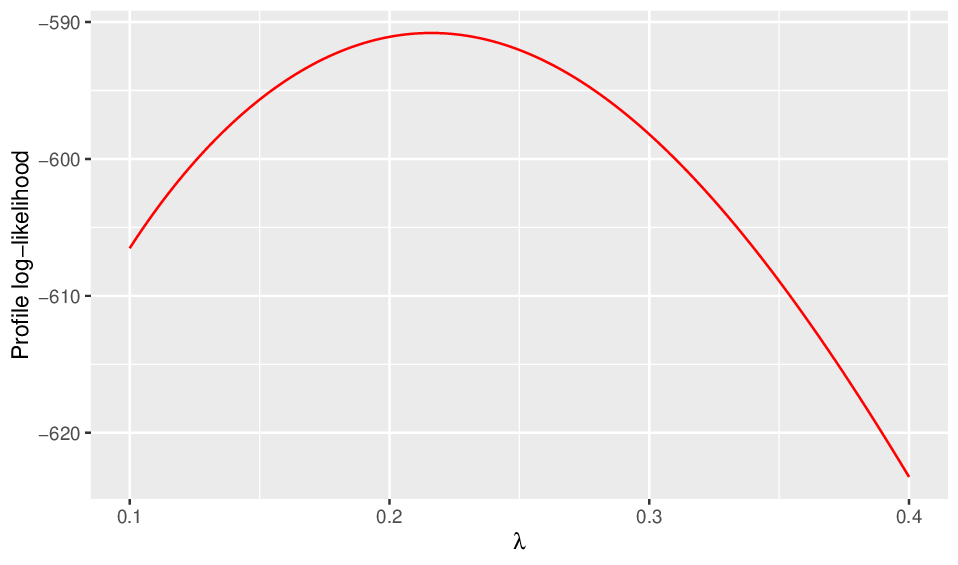}}
			\caption{ Profile log-likelihood function of  $(a)$ $\alpha$ and $(b)$ $\lambda$, for the real dataset in Section $7$. }
		\end{center}
	\end{figure}
	It is worth pointing that in statistical inference, it is always of
	interest to study the existence and uniqueness of the MLEs of the parameters. In order to achieve this, one requires
	to show two conditions proposed by \cite{makelainen1981existence}. These
	are difficult to establish due to the complicated nature of the expressions
	of the second order partial derivatives of the log-likelihood
	function. However, to have a rough idea about the existence and
	uniqueness of the MLEs of $\alpha$ and
	$\lambda$ for the LE distribution under PT-IHC sample, the profile log-likelihood
	function for the parameters are presented in Figure $2.$ These plots
	show that the MLEs may exist uniquely.  Note that the MLEs can be
obtained efficiently using numerical approximation techniques. However, the exact distribution
of the MLEs is hard to derive. As a result, the confidence intervals of the unknown model parameters are
difficult to produce. The observed Fisher information matrix can be used to construct
asymptotic confidence intervals of the parameters, which is discussed in Section $5$. Since it is not possible to derive the exact
distribution of the MLEs, the natural choice is the Bayesian inference.
This is presented in the next section.

\section{Bayesian estimation\setcounter{equation}{0}}
	This section deals with the derivation of the Bayes estimates of the
	parameters $\alpha$ and $\lambda$ of the LE
	distribution with respect to three loss functions when PT-IHC sample is available. The
	squared error, LINEX  and the generalized
	entropy loss functions are considered. Among these, the squared error loss function (SELF)
	is symmetric and other two are asymmetric. Let $\delta$ be an
	estimator of the  parameter $\eta$. Then, the SELF is given by
	\begin{eqnarray}
		L_{SQ}(\eta,\delta)= (\delta-\eta)^2 \label{eq3.1},
	\end{eqnarray}
	which assigns equal weight in
	underestimation as well as in overestimation. Sometimes, it is also
	called balanced loss function. There are various situations, where
	symmetric loss function is not an appropriate tool to use. For
	example, when estimating lifetime of a satellite, the overestimation is
	more serious than underestimation. Further, for estimating water
	level of a river in rainy season, the underestimation is more serious
	than overestimation. In these situations, the following loss
	functions, known as the LINEX loss (see \citet{Varian}) and
	generalized entropy loss (see \cite{calabria1994bayes}) are
	useful, respectively given by
	\begin{eqnarray}
		L_{LI}(\eta,\delta)&=& e^{p(\delta-\eta)}-p(\delta-\eta)-1,~~p\neq 0;\label{eq3.2}\\
		L_{GE}(\eta,\delta)&=& \left(\frac{\delta}{\eta}\right)^q
		-q\log\left({\frac{\delta}{\eta}}\right)-1,~~ q \neq 0.
		\label{eq3.3}
	\end{eqnarray}
	 Note that  the sign of $p$ in the LINEX loss function represents the direction of asymmetry, and its magnitude reflects the degree of its asymmetry. For $p<0$, a negative error has a more serious effect, and for $p>0$ the effect of a positive error is more serious. Under the loss functions given by (\ref{eq3.1}), (\ref{eq3.2}) and
	(\ref{eq3.3}), the Bayes estimates can be written in terms of the conditional
	expectations as
	\begin{eqnarray}
		\widehat{\eta}_{SQ}&=&E_{\eta}(\eta|data); \label{eq3.4}\\
		\widehat{\eta}_{LI}&=&-p^{-1}\log[E_{\eta}(e^{-p\eta}|data)],~~ p \neq 0; \label{eq3.5}\\
		\widehat{\eta}_{GE}&=&[E_{\eta}(\eta^{-q}|data)]^{-\frac{1}{q}},~~
		q \neq 0. \label{eq3.6}
	\end{eqnarray}
	In Bayesian estimation, prior distribution has an important role.
	According to \citet{arnold1983bayesian}  there is no way, in which someone can
	say which prior is better than the other. They additionally stated
	that it is all the more regularly the case that somebody chooses to
	confine consideration regarding a given adaptable group of priors,
	and then picks one from that family which appears to be the best
	match with their own convictions. In this case, two different types of priors are considered
and then, the Bayes estimates are derived.
	
	\subsection{Independent priors}
	 We recall that when both the model parameters are unknown,
 no joint conjugate prior is available. For such situation, there are number of ways to consider the priors.
 For the present study, we consider independent gamma priors. This is due to the fact that the gamma distribution has a log-concave density function in the interval $(0,\infty)$  and Jeffery's prior can be obtained as a special case of  the gamma prior. Denote by $Gamma(a,b)$ for the gamma distribution with shape parameter $a$ and scale parameter $1/b$.  Let $\alpha\sim
	Gamma(a,b)$ and $\lambda\sim Gamma(c,d)$. The PDFs of $\alpha$ and $\lambda$ are respectively given
	by
	\begin{eqnarray}
		\pi_{1}(\alpha) \propto \alpha^{a-1}
		e^{-b\alpha}~~\mbox{and}~~\pi_{2}(\lambda) \propto \lambda^{c-1}
		e^{-d\lambda}, \label{eq3.7}
	\end{eqnarray}
	where $\alpha,~\lambda>0,~~a,~b,~c,~d>0$. The joint prior distribution of $\alpha$ and $\lambda$ is
	\begin{eqnarray}
		\pi_{3}(\alpha,\lambda) \propto \alpha^{a-1} e^{-b\alpha}
		\lambda^{c-1} e^{-d\lambda};~\alpha,~\lambda>0,~~a,~b,~c,~d>0.
		\label{eq3.7*}
	\end{eqnarray}
	After some calculations, the posterior PDF
	of $\alpha$ and $\lambda$ given the observed data is
	obtained as
	\begin{eqnarray}
		\nonumber \pi(\alpha,\lambda|data)&=&  k_{1}^{-1}
		\alpha^{D+a-1} \lambda^{D+c-1} e^{-b\alpha}
		e^{-\lambda(d-\sum_{i=1}^{D}x_{i})} \prod_{i=1}^{D} (g(x_i;\lambda))^{\alpha-1}
		[1+(g(x_i;\lambda))^{\alpha}]^{-(R_{i}+2)}\\
		&~&\times [1+(g(T;\lambda))^{\alpha}]^{-R^{*}_{D}}, \label{eq3.8}
	\end{eqnarray}
	where $k_1$ is given in $(\ref{A1})$. Now, consider a function of the parameters $\alpha$ and $\lambda$,
	say $\phi(\alpha,\lambda).$ Then, from (\ref{eq3.4}), (\ref{eq3.5})
	and (\ref{eq3.6}), the forms of the Bayes estimates of
	$\phi(\alpha,\lambda)$ with respect to the squared error, LINEX and the
	generalized entropy loss functions are given by
	\begin{eqnarray}
		\widehat{\phi}_{SQ}&=& \int_{0}^{\infty}\int_{0}^{\infty} \phi(\alpha,\lambda)\pi(\alpha,\lambda|data)d\alpha d\lambda; \label{eq3.10}\\
		\widehat{\phi}_{LI}&=&-p^{-1} \log\left[\int_{0}^{\infty}\int_{0}^{\infty} \phi(\alpha,\lambda)\pi(\alpha,\lambda|data)d\alpha d\lambda\right],~~p\ne 0; \label{eq3.11}\\
		\widehat{\phi}_{GE}&=& \left[\int_{0}^{\infty}\int_{0}^{\infty}
		\left(\phi(\alpha,\lambda)\right)^{-q}\pi(\alpha,\lambda|data)d\alpha
		d\lambda\right]^{-\frac{1}{q}},~~q\ne 0, \label{eq3.12}
	\end{eqnarray}
	respectively. Note that $(\ref{eq3.12})$ reduces to
	$(\ref{eq3.10})$, for $q=-1.$ To derive the Bayes
	estimates of $\alpha$ and $\lambda$ with respect to the loss
	functions given by (\ref{eq3.1}), (\ref{eq3.2}) and (\ref{eq3.3}),
	one needs to replace $\phi(\alpha,\lambda)$ by $\alpha$ and $\lambda$,
	respectively in (\ref{eq3.10}), (\ref{eq3.11}) and (\ref{eq3.12}).
	In the following, the form of the Bayes estimates of $\alpha$ with
	respect to the loss functions (\ref{eq3.1}), (\ref{eq3.2}) and
	(\ref{eq3.3}) are respectively obtained as
	\begin{eqnarray}
		 \widehat{\alpha}_{SQ}&= &
		\int_{0}^{\infty}\int_{0}^{\infty} \alpha \pi(\alpha,\lambda|data) \ d\alpha d\lambda;\\ \label{eq3.13}
		\widehat{\alpha}_{LI}&=& -p^{-1}
		\log\Bigg[ \int_{0}^{\infty}\int_{0}^{\infty} \alpha \pi(\alpha,\lambda|data) d\alpha d\lambda\Bigg]; \\\label{eq3.14}
		\widehat{\alpha}_{GE}&=&\Bigg[ \int_{0}^{\infty}\int_{0}^{\infty} \alpha^{-q} \pi(\alpha,\lambda|data) d\alpha d\lambda\Bigg]^{-\frac{1}{q}}.
		\label{eq3.15}
	\end{eqnarray}
	In analogy to $\widehat{\alpha}_{SQ}$, $\widehat{\alpha}_{LI}$ and
	$\widehat{\alpha}_{GE}$, the Bayes estimates of $\lambda$ under the
	loss functions (\ref{eq3.1}), (\ref{eq3.2}) and (\ref{eq3.3})
 can be obtained, which are denoted by  $\widehat{\lambda}_{SQ}$, $\widehat{\lambda}_{LI}$
	and $\widehat{\lambda}_{GE}$, respectively. These are omitted for the
	sake of conciseness.
	
	\subsection{Bivariate prior}
	In this subsection, let us consider a  bivariate prior for $\alpha$
	and $\lambda$, which is given by
	\begin{eqnarray}\label{eq3.17}
		\pi^*(\alpha,\lambda)=\pi_{1}^*(\alpha|\lambda)\pi_{2}(\lambda);~~\alpha>0,~\lambda>0.
	\end{eqnarray}
	where $\pi_{2}(\lambda)$ is given by (\ref{eq3.7}) and
	\begin{eqnarray}
		\pi_{1}^*(\alpha|\lambda)\propto
		\frac{1}{\lambda};~~\alpha>0,~\lambda>0.
	\end{eqnarray}
	The prior $\pi_{1}^*(\alpha|\lambda)$ is known as the noninformative
	prior of $\alpha$ when $\lambda$ is fixed. Thus, from
	(\ref{eq3.17}), the bivariate prior distribution of $\alpha$ and
	$\lambda$ is given by
	\begin{eqnarray}
		\pi^*(\alpha,\lambda)\propto
		{\lambda}^{(c-2)}e^{-d\lambda};~~\alpha,~\lambda>0,~~c,~d>0.
	\end{eqnarray}
	Similar to the preceding subsection, the posterior PDF
	of $\alpha$ and $\lambda$ given the observed data is obtained
	as
	\begin{eqnarray}
		\pi_{2}^*(\alpha,\lambda|data)& =& k_{2}^{-1} \alpha^{D}
		\lambda^{D+c-2} e^{-\lambda(d-\sum_{i=1}^{D}x_{i})} \prod_{i=1}^{D}
		(g(x_i;\lambda))^{\alpha-1}
		[1+(g(x_i;\lambda))^{\alpha}]^{-(R_{i}+2)}\nonumber \\
		&~& \times [1+(g(T;\lambda))^{\alpha}]^{-R^{*}_{D}},
		\label{eq3.16}
	\end{eqnarray}
	where $k_2$ is given in $(\ref{A2})$. Based on the posterior density function given by (\ref{eq3.16}), the
	Bayes estimates of the parameters $\alpha$ and $\lambda$ can be
	obtained with respect to the loss functions (\ref{eq3.1}),
	(\ref{eq3.2}) and (\ref{eq3.3}). The expressions are omitted for the
	sake of brevity. Considering the bivariate prior distributions,
	the Bayes estimates of $\alpha$ under the loss functions
	(\ref{eq3.1}), (\ref{eq3.2}) and (\ref{eq3.3}) are denoted by
	$\hat{\alpha}_{dSQ}$, $\hat{\alpha}_{dLI}$ and $\hat{\alpha}_{dGE}$,
	respectively. Further, the Bayes estimates of $\lambda$ under the
	loss functions (\ref{eq3.1}), (\ref{eq3.2}) and (\ref{eq3.3}) are
	denoted by $\hat{\lambda}_{dSQ}$, $\hat{\lambda}_{dLI}$ and
	$\hat{\lambda}_{dGE}$, respectively.\\

	\section{Approximation techniques}
	There are various techniques in the literature which have been used
	to compute approximate Bayes estimates. In this section, two
	approaches are employed for the evaluation of the Bayes estimates
	obtained in the preceding section. First, consider Lindley's method,
	which expresses how to approximate a ratio of two particular forms of
	integrations. For details, please refer to \cite{Lindley}.
	\subsection{Lindley's approximation method}
	This subsection deals with the computation of the Bayes estimates of
	$\alpha$ and $\lambda$ of the LE distribution under
	the loss functions given by (\ref{eq3.1}), (\ref{eq3.2}) and (\ref{eq3.3})
	when PT-IHC sample is available.
	First, let us write the Bayes estimate of $\alpha$ with respect to
	the LINEX loss function. Please refer to Appendix B for the idea on Lindley's
	approximation method. In this case, for $p\ne 0,$ one has
	$$\phi(\alpha,\lambda)=e^{-p\alpha},~ v_{1}=-pe^{-p\alpha},~v_{11}=p^2e^{-p\alpha},~ v_{2}=v_{12}=v_{21}=v_{22}=0.$$
	Thus, using $(\ref{a1})$, the approximate Bayes estimate of $\alpha$
	under LINEX loss function is obtained as
	\begin{eqnarray}
		\nonumber \widehat{\alpha}_{LI}&\approx&  -\left(\frac{1}{p}\right)\log\Big[ e^{-p\alpha}+\frac{1}{2}\{p^2e^{-p\alpha}-pe^{-p\alpha}( l_{30}\tau^{2}_{11} + l_{03}\tau_{21}\tau_{22} + 3l_{21}\tau_{11}\tau_{12}\nonumber\\
		&~&+
		l_{12}(\tau_{11}\tau_{22}+2\tau^2_{21})+2P_{1}\tau_{11}+2P_{2}\tau_{21})
		\} \Big], \label{eq4.1}
	\end{eqnarray}
	where the other unknown terms in (\ref{eq4.1}) are given in Appendix
	B. Next, consider the general entropy loss function given by
	(\ref{eq3.3}). Under this loss function,
	$$\phi(\alpha,\lambda)=\alpha^{-q},~ v_{1}=-q\alpha^{-(q+1)},~ v_{2}=q(q+1)\alpha^{-(q+2)},~ v_{2}=v_{12}=v_{21}=v_{22}=0.$$ Therefore, under the general entropy loss function, the approximate Bayes estimate of $\alpha$ can be obtained, which is given by
	\begin{eqnarray}
		\nonumber \widehat{\alpha}_{GE}&\approx&\Big[\alpha^{-q}+0.5\{q(q+1)\alpha^{-(q+2)}-q\alpha^{-(q+1)}( l_{30}\tau^{2}_{11} + l_{03}\tau_{21}\tau_{22} + 3l_{21}\tau_{11}\tau_{12}\\
		&~& +
		l_{12}(\tau_{11}\tau_{22}+2\tau^2_{21})+2P_{1}\tau_{11}+2P_{2}\tau_{21})
		\}\Big ]^{-\frac{1}{q}}.\label{eq4.18}
	\end{eqnarray}
	Note that the Bayes estimate of $\alpha$ with respect to the SELF (denoted by $\widehat{\alpha}_{SQ}$) can be computed
	from (\ref{eq4.18}) when $q=-1.$ The Bayes estimates of
	$\lambda$ with respect to the LINEX, generalized entropy and the squared
	error loss functions can be derived similarly, and
	therefore the expressions are omitted. The Bayes estimates of
	$\lambda$ with respect to the loss functions (\ref{eq3.1}),
	(\ref{eq3.2}) and (\ref{eq3.3}) using Lindley's approximation
	technique are denoted by $\widehat{\lambda}_{SQ}$, $\widehat{\lambda}_{LI}$
	and $\widehat{\lambda}_{GE},$ respectively. Further, when the bivariate
	prior distribution for $\alpha$ and $\lambda$ (see Subsection $3.2$)
	is considered, the approximate Bayes estimates for $\alpha$ and
	$\lambda$ can be obtained similar to the case of independent priors.
	Some changes in computing $P_1$ and $P_2$ are required in this
	purpose.

	\subsection{Importance sampling method}
	In this subsection, importance sampling method is used to evaluate
	the Bayes estimates of the parameters $\alpha$ and $\lambda$ with
	respect to the squared error, LINEX and the generalized entropy loss functions.
	The main advantage of this method over the Lindley's approximation
	method is that this method can be used to
	construct HPD credible intervals. In
	this method, one needs to rewrite the posterior PDF given by Equation (\ref{eq3.8}) as
	
	\begin{eqnarray}
		\pi(\alpha,\lambda|data) \propto
		G_{\lambda}\left(D+c,d-\sum_{i=1}^{D}x_{i}\right)G_{\alpha|\lambda}\left(D+a,b-\sum_{i=1}^{D}\log
		(g(x_i;\lambda)) \right) h(\alpha,\lambda), \label{eq5.1*}
	\end{eqnarray}
	where $G(a_1,a_2)$ represents gamma distribution with parameters
	$a_1$ and $a_2$, and $h(\alpha,\lambda)$ is given in $(\ref{A6})$.
	
	Further, in Eq. (\ref{eq5.1*}),
	$G_{\lambda}\left(D+c,d-\sum_{i=1}^{D}x_{i}\right)$ and
	$G_{\alpha|\lambda}\left(D+a,b-\sum_{i=1}^{D}\log (g(x_i;\lambda)) \right)$ mean $$\lambda\sim
	G\left(D+c,d-\sum_{i=1}^{D}x_{i}\right)~~ \mbox{and}~~
	\alpha|\lambda \sim G\left(D+a,b-\sum_{i=1}^{D}\log (g(x_i;\lambda)) \right).$$ Now, the algorithm comprising of the following
	steps can be used to obtain Bayes estimates of
	$\phi(\alpha,\lambda)$ with respect to the loss functions given by
	(\ref{eq3.1}), (\ref{eq3.2}) and (\ref{eq3.3}).
	\\
	 ------------------------------------------------------------------------------------------------------------------\\
	Algorithm-1\\
	 ------------------------------------------------------------------------------------------------------------------
	\begin{itemize}
		\item[Step-1:] Generate $\lambda$ from  $G\left(D+c,d-\sum_{i=1}^{D}x_{i}\right)$ distribution.
		
		\item[Step-2:] For the given value of $\lambda$ in Step-$1$, generate $\alpha$ from $G\left(D+a,b-\sum_{i=1}^{D}\log (g(x_i;\lambda)) \right)$ distribution.
		
		\item[Step-3:] Repeat Step-1 and Step-2, for $N$ times to obtain $(\alpha_{1},\lambda_{1}),\ldots,(\alpha_{N},\lambda_{N}).$
	\end{itemize}
	 -------------------------------------------------------------------------------------------------------------------\\
	Based on the values as in Step-3, the Bayes estimates of a
	parametric function $\phi(\alpha,\lambda)$ under the LINEX and the
	generalized entropy loss functions are obtained as
	\begin{eqnarray}
		\tilde{\phi}_{LI}=-\left(\frac{1}{p}\right) \log
		\left[\frac{\sum_{i=1}^{N}e^{-p\phi(\alpha_{i},\lambda_{i})}h(\alpha_{i},\lambda_{i})}{\sum_{i=1}^{N}
			h(\alpha_{i},\lambda_{i})}\right]\label{eq5.3}
	\end{eqnarray}
	and
	\begin{eqnarray}
		\tilde{\phi}_{GE}=
		\left[\frac{\sum_{i=1}^{N}\phi^{-q}(\alpha_{i},\lambda_{i})h(\alpha_{i},\lambda_{i})}
		{\sum_{i=1}^{N}h(\alpha_{i},\lambda_{i})}\right]^{-\frac{1}{q}},
		\label{eq5.4}
	\end{eqnarray}
	respectively.  The Bayes estimate with respect to the SELF, denoted by $\tilde{\phi}_{SQ}$ can be obtained from
	(\ref{eq5.4}) when $q=-1.$ Further, the Bayes estimates of $\alpha$
	with respect to the loss functions (\ref{eq3.2}) and (\ref{eq3.3})
	can be  respectively obtained from (\ref{eq5.3}) and (\ref{eq5.4})
	after substituting $\alpha_{i}$ in place of
	$\phi(\alpha_{i},\lambda_{i})$. Similarly, the Bayes
	estimates of $\lambda$ can be evaluated. The Bayes estimates of $\alpha$ (respectively $\lambda$)
	with respect to the loss functions given by  (\ref{eq3.1}),
	(\ref{eq3.2}) and (\ref{eq3.3}) obtained using importance sampling
	method are denoted by $\tilde{\alpha}_{SQ}$
	(respectively  $\tilde{\lambda}_{SQ}$), $\tilde{\alpha}_{LI}$
	(respectively  $\tilde{\lambda}_{LI}$) and $\tilde{\alpha}_{GE}$
	(respectively  $\tilde{\lambda}_{GE}$), respectively. It is known that the
	credible intervals are constructed based on the posterior
	distribution. Further, if the credible interval satisfies the
	condition of HPD interval, then it is called as the HPD credible
	interval. In this paper, the $100(1-\beta)\%$ HPD credible intervals of
	the parameters $\alpha$ and $\lambda$ are constructed numerically
	based on the generated posterior samples. In this context, the idea
	proposed by \citet{chen1999monte}  has been used. Note that although the importance sampling method is used to compute Bayes estimates and then HPD credible intervals, the Bayes estimates using importance sampling method are not presented in Section $6$ for brevity.
	
	\section{Interval estimation}\setcounter{equation}{0}
	In this section, two methods are employed to construct
	confidence intervals of $\alpha$ and $\lambda.$ Using the asymptotic
	normality property of the MLEs
	$\widehat{\alpha}$ and $\widehat{\lambda}$, the $100(1-\beta)\%$ confidence
	intervals of the unknown model parameters $\alpha$ and $\lambda$ can
	be constructed. In doing so, the variances of   $\widehat{\alpha}$ and
	$\widehat{\lambda}$ are required, which can be obtained from the inverse
	of the observed Fisher information matrix. Note that under some
	regularity conditions, the MLEs
	$(\widehat{\alpha},\widehat{\lambda})$ asymptotically follow  bivariate
	normal distribution with mean vector $(\alpha,\lambda)$ and
	variance-covariance matrix $I^{-1}(\widehat{\alpha},\widehat{\lambda})$.
	That is,
	$$(\widehat{\alpha},\widehat{\lambda})\sim N((\alpha,\lambda),\widehat{I}^{-1}(\widehat{\alpha},\widehat{\lambda})),$$
	where $\widehat{I}(\widehat{\alpha},\widehat{\lambda})$ is the observed Fisher
	information matrix given by
	\begin{align}\label{eq5.1}
		\widehat{I}(\widehat{\alpha},\widehat{\lambda})= {\begin{bmatrix}
				-l_{20} & -l_{11}\\
				-l_{11} & -l_{02}\\
		\end{bmatrix}}_{({\alpha},{\lambda})=(\widehat{\alpha},\widehat{\lambda})}.
	\end{align}
	The elements of the observed Fisher information matrix given by
	(\ref{eq5.1}) are provided in Appendix B. Note that the variances
	of $\widehat{\alpha}$ and $\widehat{\lambda}$, respectively denoted by
	$Var(\widehat{\alpha})$ and  $Var(\widehat{\lambda})$ can be obtained from
	the main diagonal entries of the matrix
	$\widehat{I}^{-1}(\widehat{\alpha},\widehat{\lambda}).$ Thus, the
	$100(1-\beta)\%$ approximate confidence intervals for $\alpha$ and
	$\lambda$ are constructed as
	$$\left(\widehat{\alpha} \pm z_{\frac{\beta}{2}}\sqrt{Var(\widehat{\alpha})}\right)~~\mbox{ and}~~  \left(\widehat{\lambda} \pm z_{\frac{\beta}{2}}\sqrt{Var(\widehat{\lambda})}\right),$$
	where $z_{\beta/2}$ is the percentile of the standard normal
	distribution with right-tail probability $\beta/2$. In some cases,
	the $100(1-\beta)\%$ confidence interval has negative lower bound
	when the parameter takes positive value. This is a drawback of the
	normal approximation of the MLEs. To overcome
	this drawback, \cite{meeker2014statistical} introduced log-transformed MLE-based confidence intervals. According to
	these authors, the confidence interval obtained based on logarithm
	transformed MLEs has better coverage probability. Based on the
	log-transformed MLEs, the $100(1-\beta)\%$ confidence intervals of
	$\alpha$ and $\lambda$ are respectively given by
	$$\left(\widehat{\alpha} \exp\left\{\pm\frac{z_{\frac{\beta}{2}}\sqrt{\tau_{11}}}{\widehat{\alpha}}\right\}\right)~~ \mbox{and}~~ \left(\widehat{\lambda} \exp\left\{\pm\frac{z_{\frac{\beta}{2}}\sqrt{\tau_{22}}}{\widehat{\lambda}}\right\}\right).$$
	
	\section{A simulation study}\setcounter{equation}{0}
	In this section, a Monte Carlo simulation is performed to
	investigate the performance of the proposed estimates for LE distribution under PT-IHC sample. The comparative performance of the point estimates
	is studied in terms of the average and mean squared error (MSE)
	values. The interval estimates are compared based on the average lengths.  In doing so, $2000$ PT-IHC
	samples are generated using various censoring schemes. Several combinations of $(n,m)$ are considered. The values of $T$ are taken
	as $0.5$ and $0.65$. The $R~3.4.0$ software is used for the
	purpose of simulation. Two different (total) sample sizes $n=35$ and $40$
	are considered here. The numerical study includes the following
	steps:
	\begin{enumerate}
		\item  True values of the model parameters are assumed as $\alpha= 1.5$ and $\lambda=0.75$.
		\item The PT-IHC samples for different $n$ and $m$ have been generated. Here, the values of $(n,m)$ are considered as $(35,10)$, $(35,25)$, $(40,10)$ and $(40,30)$.
		\item In the tables, ($0*3$) and ($1*5$) represent $(0,0,0)$ and $(1,1,1,1,1)$, respectively.
		For each censored schemes, average values of the MLEs and the Bayes estimates corresponding to the univariate and bivariate priors and the associated MSEs are tabulated  in Tables \ref{Table 1}, \ref{Table 2},
		\ref{Table 5} and \ref{Table 6}. Note that the comparisons of the proposed estimates are made based on the average and MSE values. To compute MSE, the following formula is used:
		\begin{eqnarray}
			\nonumber MSE= \frac{1}{M} \sum_{i=1}^{M}
			\left[\theta^{(i)}_{k}-\theta_{k}\right]^2;  & i=1,~2,
		\end{eqnarray}
		where $\theta_{1}=\alpha$, $\theta_{2}=\lambda$ and $M=2000$.
		
		\item  The Bayes estimates of $\alpha$ and $\lambda$ are obtained with respect to the squared error loss function (SELF), LINEX loss function (LLF) and the generalized entropy loss function (GELF). In this purpose, the hyperparameters  are considered as  $a=3,~b=2,~c=3$ and $d=4$. We note that Lindley's approximation method is used to compte the estimated values of the Bayes estimates. To obtain Bayes estimates under the LLF and the GELF,
		three choices of $p$ and $q$ are considered as $p=(-0.05,0.5,1)$ and $q=(-0.5,-0.25,0.25)$.
		
		\item  In Tables \ref{Table 3} and \ref{Table 7}, the computed average lengths of $90\%$ and $95\%$ confidence intervals  using normal approximation to MLEs (NA) and normal approximation to the log-transform MLEs (NL) methods are
		presented for $\alpha$ and $\lambda$. The average lengths of the HPD credible intervals of these parameters are also provided there. It is worth mentioning that the importance sampling method is utilized in order to get the HPD credible intervals since Lindley's approximation technique can not be used to obtain the same.
			\end{enumerate}
		In this part of the section, the coverage probabilities of the
		unknown parameters are discussed. To this aim, the asymptotic
		pivotal quantities are defined as
		\begin{align}
			\nonumber Q_{1}=\frac{\widehat{\alpha}-\alpha}{\widehat{\lambda}\sqrt{\tau_{11}}},
			& &  Q_{2}= \frac{\widehat{\alpha}-\alpha}{{\lambda}\sqrt{\tau_{11}}},
			& &  Q_{3}= \frac{\widehat{\lambda}-\lambda}{\widehat{\lambda}\sqrt{\tau_{22}}}.
		\end{align}
		By using Monte Carlo simulation, the coverage probabilities
		\begin{align}
			\nonumber P(-1.65 \leq Q_{i} \leq 1.65) ~~\mbox{and} ~~  P(-1.96 \leq Q_{i} \leq 1.96)
		\end{align}
		for $i=1,~2,~3$ are computed, which are approximately $90\%$ and $95\%$, respectively (see Tables $4$ and $8$). From Tables $1$-$8$, the following observations have been made.
	\begin{enumerate}
		\item Under the same censoring schemes when $T$ increases,  MSE decreases. Further, for fixed $n$, when $m$ increases, then MSE decreases.
		
		\item For the parameter $\alpha$, in general, the Bayes estimates under GELF perform better than other estimates, whereas, for $\lambda$, most of the times, the Bayes estimates under LLF perform better than other estimates. To estimate $\alpha$, LLF performs better when $p=1$, and GELF perform better when $q=0.25$ and to estimate $\lambda$, LLF performs better when $p=-0.05$, and GELF performs better when $q=-0.5$. The Bayes estimates under SELF are almost equal to the value of the Bayes estimates under LLF when $p$ is nearly $0$. Further, the Bayes estimates under the univariate (U) prior perform better than the estimates with respect to the bivariate (B) priors in most of the cases.
		
		\item When $T$ increases, length of the confidence intervals decrease. For each censoring schemes, the lengths of the NA-based confidence intervals are smaller than that of the NL-based confidence intervals.
		\item The HPD credible interval lengths are smaller than other simulated interval lengths. Hence, HPD credible intervals perform better than other intervals.
		\item  The coverage probabilities of the above mentioned pivotal quantities $Q_1$, $Q_2$ and $Q_3$ increase
		when $m$ and $T$ both increase. Furthermore, coverage probabilities under $Q_{2}$ are much better than other two pivotal quantities.
	\end{enumerate}

	\begin{table}[!htbp]
		\begin{center}
		\renewcommand\thetable{1}
		\scriptsize \caption{\label{Table 1}Average and MSE values of the
			estimates of $\alpha$ (true value=1.5) for $T=0.5$.  The first (second) rows corresponding to `U' and `B' are for the average (MSE) values of the estimates.}
		\scalebox{0.9}{
			\begin{tabular}{cccccccccccccccccc}
				\hline\\
				(n,m)& scheme & prior &$\widehat{\alpha}$&  &$\widehat{\alpha}_{LI}$&  & &$\widehat{\alpha}_{GE}$     & && $\widehat{\alpha}_{SQ}$ \\
				\hline\\
				& & & &$p=-0.05$ & $p=0.5$ & $p=1$  & $q=-0.5$ & $q=-0.25$ & $q=0.25$ && \\
				\hline \\
				(35,10)& (0*9,25)&U& 1.7716 & 1.6524 & 1.5913 &1.5467 &1.6142& 1.5993& 1.5721& &1.6464 \\
				&  && 0.0737 & 0.0232 & 0.0083 & 0.0022 &0.0131& 0.0099& 0.0052& & 0.0214\\
				& &  B& & 1.6626&1.5981 &1.5483 & 1.6219&1.6054 &1.5748 & & 1.6565 \\
				& & & & 0.0264& 0.0096& 0.0023& 0.0148& 0.0112& 0.0056& & 0.0245\\ [0.1 cm]
				& (0*5,5*5)& U &1.7555& 1.5789& 1.5335& 1.5021& 1.5502& 1.5393& 1.5199& & 1.5744\\
				&  && 0.0653 & 0.0062& 0.0011& 0.0004&  0.0025& 0.0015& 0.0004& & 0.0055\\
				& & B& & 1.6092& 1.5368&1.4785 & 1.5716&1.5374 & 1.5032& & 1.5937 \\
				& & & & 0.0119& 0.0014& 0.0005& 0.0051& 0.0014& 0.0001& & 0.0087\\ [0.1 cm]
				& (25,0*9)& U& 1.7059& 1.5201& 1.4763& 1.4464&  1.4917& 1.4811& 1.4623& & 1.5156\\
				& & & 0.0424&  0.0004& 0.0006& 0.0028&  0.0001& 0.0004& 0.0014& & 0.0002\\
				& & B& & 1.6431& 1.5750 &1.5239 & 1.5999&1.5828 & 1.5525& & 1.6365 \\
				& & & & 0.0204& 0.0056& 0.0006& 0.0099&0.0068& 0.0027& & 0.0186\\
				\hline\\
				(35,25)& (0*24,10)&U& 1.7217& 1.5265& 1.4764& 1.4429 & 1.4942& 1.4451&  1.4231& & 1.4874  \\
				& &  &0.0491&  0.0007& 0.0005& 0.0032&  0.0000& 0.0030& 0.0059& & 0.0001   \\
				& &B& &1.6320&1.5751 &1.5302 & 1.5933&1.5806 &1.5527 & & 1.6266 \\
				& & & & 0.0174& 0.0056& 0.0014& 0.0087& 0.0065& 0.0027& & 0.0160\\ [0.1 cm]
				& (0*20,2*5)&U&  1.6923& 1.4860& 1.4331& 1.4248&   1.4748& 1.4628& 1.4419& & 1.5015 \\
				& & &0.0369& 0.0002& 0.0044& 0.0056&  0.0006&  0.0013& 0.0033& & 0.0001     \\
				& & B& & 1.6410&1.5833 &1.5380 & 1.6041&1.5892 &1.5611 & & 1.6355 \\
				& & & & 0.0198& 0.0069& 0.0009& 0.0108& 0.0079& 0.0037& & 0.0183 \\ [0.1 cm]
				& (10,0*24)&U&  1.6781& 1.5209& 1.4728& 1.4388&  1.4894&  1.4773& 1.4558& & 1.5161 \\
				& &  &0.0317&  0.0004&  0.0007&   0.0037&  0.0001&  0.0005& 0.0019& & 0.0003 \\
				& & B& & 1.6314&1.5848 &1.5472 & 1.6015&1.5892 &1.5658 & & 1.6270\\
				& & & & 0.0172& 0.0072& 0.0022& 0.0103& 0.0079& 0.0043& & 0.0161 \\
				\hline\\
				(40,10)& (0*9,30)&U&  1.7104& 1.5410& 1.4965& 1.4653&  1.5123& 1.5013& 1.4819& & 1.5365  \\
				& &   &0.0442& 0.0016& 0.0001& 0.0011&   0.0002& 0.0001& 0.0003& & 0.0013  \\
				& & B& & 1.6474&1.5815 &1.5312 & 1.6055& 1.5888 &1.5577 && 1.6410\\
				& & & & 0.0217& 0.0066& 0.0009& 0.0111& 0.0078& 0.0033& & 0.0199 \\ [0.1 cm]
				& (0*5,6*5)&U&  1.7585& 1.5684&  1.5252& 1.4958 &  1.5410& 1.5308&  1.5127& & 1.5640 \\
				& &  &0.0668& 0.0046& 0.0006 &0.0001& 0.0017& 0.0009& 0.0002& & 0.0041  \\
				& & B& & 1.7007&1.6387 &1.5895 & 1.6620&1.6462 &1.6164 & & 1.6948\\
				& & & & 0.0402& 0.0192& 0.0080& 0.0262& 0.0213& 0.0135& & 0.0379 \\ [0.1 cm]
				& (30,0*9)&U& 1.6770& 1.6091&  1.5394& 1.4860& 1.5639& 1.5457& 1.5121& &  1.6025 \\
				& &  &0.0313&  0.0119& 0.0015& 0.0002& 0.0041& 0.0021& 0.0002& & 0.0105  \\
				& & B& & 1.6618&1.6078 &1.5716 & 1.6269&1.6121 &1.5835 & & 1.6564 \\
				& & & & 0.0262& 0.0116& 0.0051& 0.0161& 0.0125& 0.0069& & 0.0244 \\
				\hline\\
				(40,30)& (0*29,10)&U& 1.6558& 1.4926& 1.4455& 1.4124& 1.4614& 1.4495& 1.4284& & 1.4879 \\
				& & &0.0242&  0.0001& 0.0029& 0.0077&  0.0015& 0.0025& 0.0051& & 0.0001  \\
				& & B& & 1.6117&1.5634 &1.5242 & 1.5802& 1.5673&1.5427 & & 1.6071 \\
				& & & & 0.0124& 0.0040& 0.0005& 0.0064& 0.0045& 0.0018& & 0.0114 \\ [0.1 cm]
				& (20*0,1*10)&U&  1.6869& 1.5367& 1.4919& 1.4599&  1.5076&  1.4963&  1.4762& &  1.5322 \\
				& & &  0.0349&  0.0013& 0.0001& 0.0016&  0.0001&0.0000&  0.0005& & 0.0010  \\
				& & B& & 1.6009&1.5507 &1.5115 & 1.5684&1.5552 &1.5307 & & 1.5961 \\
				& & & & 0.0101& 0.0025& 0.0001& 0.0046& 0.0031& 0.0009& & 0.0092 \\ [0.1 cm]
				& (10,0*29)& U& 1.6670& 1.5330&  1.4881& 1.4552&  1.5036& 1.4921&  1.4713& &  1.5286 \\
				& &  &0.0279& 0.0011& 0.0001&  0.0020&  0.0000& 0.0001& 0.0008& & 0.0008 \\
				& & B& & 1.5887&1.5494 &1.5169 & 1.5628&1.5521 &1.5315 & & 1.5851 \\
				& & & & 0.0078& 0.0024& 0.0003& 0.0039& 0.0027& 0.0009& & 0.0072 \\
				\hline
			\end{tabular}
		}
	\end{center}
	\end{table}

	\begin{table}[!htbp]
		\begin{center}
		\renewcommand\thetable{2}
		\scriptsize \caption{\label{Table 2}Average and MSE values of the
			estimates of $\lambda$ (true value=0.75) for $T=0.5$.  The first (second) rows corresponding to `U' and `B' are for the average (MSE) values of the estimates. }
		\scalebox{0.9}{
			\begin{tabular}{cccccccccccccccccc}
				\hline\\
				(n,m)& scheme &prior & $\widehat{\lambda}$&  &$\widehat{\lambda}_{LI}$&   &  &$\widehat{\lambda}_{GE}$     & && $\widehat{\lambda}_{SQ}$      \\
				\hline\\
				& & && $p=-0.05$ & $p=0.5$ & $p=1$ &  $q=-0.5$ & $q=-0.25$ & $q=0.25$ && \\
				\hline\\
				(35,10)& (0*9,25)&U& 0.7970 & 0.7549 &0.7457 &0.7379&  0.7437& 0.7388& 0.7297 & & 0.7540   \\
				& & & 0.0022& 0.0001& 0.0001 &0.0001& 0.0001& 0.0001& 0.0004& &0.0001\\
				& & B& & 0.6975&0.6900 &0.6840 &0.6885&0.6849 &0.6785 & & 0.6968 \\
				& & & & 0.0027& 0.0035& 0.0043&0.0037& 0.0042& 0.0051& &0.0028\\ [0.1 cm]
				& (0*5,5*5)& U& 0.8112 & 0.7328& 0.7255& 0.7195&  0.7242 & 0.7206&  0.7141& & 0.7322\\
				&  & &0.0037 & 0.0003& 0.0006& 0.0009 & 0.0007&  0.0009& 0.0013& & 0.0003\\
				& & B& & 0.7339& 0.7265& 0.7197& 0.7251&0.7213 &0.7130 & & 0.7333\\
				& & & & 0.0003& 0.0005& 0.0009&0.0006&0.0008& 0.0014& & 0.0003 \\ [0.1 cm]
				& (25,0*9)&U&  0.7933& 0.7912& 0.7770& 0.7643&  0.7736& 0.7657& 0.7504& & 0.7899  \\
				& & &0.0018&  0.0017&  0.0007&  0.0002& 0.0005& 0.0002& 0.0000& & 0.0015\\
				& & B& & 0.6521&0.6429 &0.6257 & 0.6529&0.6400 &0.6211 & & 0.6512\\
				& & & & 0.0095& 0.0114& 0.0154& 0.0094& 0.0119& 0.0166& & 0.0097\\
				\hline\\
				(35,25)& (0*24,10)&U&  0.8133& 0.71966& 0.7124&  0.7066& 0.7112& 0.7274& 0.7198& &  0.7406 \\
				& &  &0.0040& 0.0009& 0.0014&  0.0018& 0.0015& 0.0005& 0.0009& &  0.0001 \\
				& & B& & 0.6814&0.6738 &0.6677 & 0.6720&0.6681 &0.6612 & & 0.6806 \\
				& & & & 0.0047& 0.0057& 0.0067& 0.0060& 0.0067& 0.0078& & 0.0048 \\ [0.1 cm]
				& (0*20,2*5)&U& 0.8102&  0.7400& 0.7316& 0.7147& 0.7194&  0.7158&  0.7092& & 0.7275   \\
				& & &0.0036&  0.0001& 0.0003&  0.0012&  0.0009& 0.0011 & 0.0016& & 0.0005    \\
				& &B& & 0.6844&0.6770 &0.6709 & 0.6752&0.6714 &0.6677 & & 0.6837 \\
				& & & & 0.0049& 0.0053& 0.0062& 0.0055& 0.0065& 0.0067& & 0.0043 \\ [0.1 cm]
				& (10,0*24)&U& 0.8125& 0.7296& 0.7202& 0.7126&  0.7185&  0.7139& 0.7057&  & 0.7287\\
				& & &0.0039&  0.0004& 0.0008& 0.0014& 0.0009& 0.0013&  0.0019& &  0.0005 \\
				& & B& & 0.6986&0.6912 &0.6852 & 0.6894&0.6857 &0.6788 & & 0.6979 \\
				& & & & 0.0026& 0.0034& 0.0041& 0.0036& 0.0041& 0.0051& & 0.0027\\
				\hline\\
				(40,10)& (0*9,30)&U& 0.8455& 0.7857& 0.7786& 0.7726& 0.7775& 0.7740&  0.7674& & 0.7851   \\
				& & &  0.0091& 0.0013& 0.0008& 0.0005& 0.0008& 0.0006& 0.0003& & 0.0012 \\
				& & B& & 0.7129&0.7054 & 0.6994& 0.7041&0.7005 &0.6941 & & 0.7122 \\
				& & & & 0.0013& 0.0019& 0.0025& 0.0021& 0.0024& 0.0031& & 0.0014 \\ [0.1 cm]
				& (0*5,6*5)&U& 0.8661& 0.7732& 0.7662& 0.7605& 0.7654& 0.7622& 0.7565& & 0.7725 \\
				& & &0.0134& 0.0005& 0.0002& 0.0001& 0.0002& 0.0001& 0.0000& & 0.0005   \\
				& & B& & 0.7470&0.7401 &0.7345 & 0.7390&0.7356 &0.7294 & & 0.7464 \\
				& & & & 0.0000& 0.0001& 0.0002& 0.0001& 0.0020& 0.0006& & 0.0000 \\ [0.1 cm]
				& (30,0*9)&U&  0.8559& 0.7376& 0.7140& 0.6958& 0.7111& 0.7007& 0.6835& & 0.7353   \\
				& & &  0.0112& 0.0001& 0.0012& 0.0029& 0.0015& 0.0024& 0.0044& & 0.0002  \\
				& & B& & 0.7230& 0.7153&0.7116 & 0.7171&0.7161 &0.7163 & & 0.7221 \\
				& & & & 0.0007& 0.0012& 0.0014& 0.0010& 0.0011& 0.0011& & 0.0007 \\
				\hline\\
				(40,30)& (0*29,10)&U& 0.8068&  0.7302& 0.7231& 0.7172&  0.7216& 0.7181& 0.7116 & & 0.7295 \\
				& & & 0.0032&  0.0004& 0.0007& 0.0011&  0.0008& 0.0010& 0.0014& & 0.0004 \\
				& & B& & 0.7510& 0.7442&0.7386 & 0.7431& 0.7397&0.7337 & & 0.7513 \\
				& & & & 0.0000& 0.0000& 0.0001& 0.0001& 0.0001& 0.0003& & 0.0000\\ [0.1 cm]
				& (20*0,1*10)&U&  0.8118& 0.7391& 0.7320& 0.7261& 0.7306& 0.7271& 0.7206& &  0.7385\\
				& & &  0.0038& 0.0001& 0.0003&  0.0005&  0.0004& 0.0005& 0.0009& & 0.0001 \\
				& & B& & 0.7275&0.7204 &0.7146 & 0.7191&0.7156 &0.7094 & & 0.7268 \\
				& & & & 0.0005& 0.0009& 0.0013& 0.0010& 0.0012& 0.0016& & 0.0005\\ [0.1 cm]
				& (10,0*29)&I&  0.8134& 0.7462& 0.7371& 0.7296& 0.7354& 0.7308& 0.7225& & 0.7453 \\
				& & & 0.0040& 0.0000& 0.0002&  0.0004& 0.0002& 0.0003&  0.0008& & 0.0000 \\
				& & B& & 0.7350&0.7277 &0.7217 & 0.7263&0.7227 &0.7162 & & 0.7343 \\
				& & & & 0.0002& 0.0004& 0.0008& 0.0006& 0.0007& 0.0011& & 0.0002 \\
				\hline\\
		\end{tabular}}
	\end{center}
	\end{table}

	\begin{table}[!htbp]
		\renewcommand\thetable{3}
		\scriptsize \caption{\label{Table 3}Average lengths of the interval
			estimates of $\alpha$ and $\lambda$ for $T=0.5$. The first (second) row corresponding to the scheme is for  $\alpha$ ($\lambda$).  The true values of $\alpha$ and $\lambda$ are $1.5$ and $0.75$, respectively.  }
		\begin{center}
		\scalebox{0.9}{
			\begin{tabular}{cccccccccccccccccc}
				\hline\\
				& &  & 90$\%$ confidence interval&  & & &&95$\%$ confidence interval \\
				\hline\\
				(n,m)& scheme& NA & NL&  HPD& & &NA & NL& HPD\\
				\hline\\
				(35,10)& (0*9,25)& 1.6629& 1.7246& 1.3470&& & 1.9753& 2.0792& 1.5329\\
				&  & 0.6246& 0.6407&0.5315 && & 0.7419& 0.7690&  0.6323 \\
				& (0*5,5*5)& 1.5285& 1.5773& 1.1250&&  & 1.8157& 1.8977& 1.1319 \\
				& & 0.6075& 0.6218& 0.5775 && & 0.7216& 0.7456& 0.6748  \\
				& (25,0*9)&  1.5198& 1.5706&1.1631 && & 1.8053& 1.8908& 1.2984  \\
				& & 0.7529& 0.7815& 0.7150&& & 0.8943& 0.9425& 0.8518    \\
				\hline\\
				(35,25)& (0*24,10)&  1.7087& 1.7809 & 1.2966 && & 2.0298& 2.1514&  1.3996\\
				& &  0.6255& 0.6412& 0.5175 && & 0.7431& 0.7693&  0.6094   \\
				& (0*20,2*5)&  1.5845& 1.6432& 1.2424& & & 1.8822& 1.9810& 1.4469     \\
				& & 0.6177& 0.6327& 0.5183& & & 0.7337& 0.7590& 0.6251    \\
				& (10,0*24)&  1.5361& 1.5903& 1.3799& & & 1.8247& 1.9160& 1.5696  \\
				& &  0.6809&  0.7010& 0.6019 & & & 0.8088& 0.8426& 0.7177  \\
				\hline\\
				(40,10)& (0*9,30)&  1.5035& 1.5523& 1.3239& & & 1.7322& 1.7859& 1.4986   \\
				& &  0.5744& 0.5855& 0.5431& & & 0.6823& 0.7010 &0.62844  \\
				& (0*5,6*5)& 1.5186&  1.5662& 1.4786& & & 1.8039& 1.8840& 1.7528\\
				& &  0.6216& 0.6350& 0.5932& & &  0.7383& 0.7609& 0.6611   \\
				& (30,0*9)& 1.7205&  1.7970& 1.6800& & & 2.0438& 2.1726& 1.9801   \\
				& & 0.8556& 0.8970& 0.8001& & & 1.0163& 1.0861& 0.9904    \\
				\hline\\
				(40,30)& (0*29,10)& 1.5315& 1.5867& 1.3237&  &&  1.8193 & 1.9122&  1.7741   \\
				& &  0.5990& 0.6129& 0.4853& &&  0.7116& 0.7349&   0.6654   \\
				& (20*0,1*10)& 1.4769&  1.5246& 1.4266 && & 1.7544& 1.8346& 1.7003   \\
				& &  0.5935& 0.6068& 0.5504 && & 0.7050&  0.7274& 0.6538 \\
				& (10,0*29)& 1.4571& 1.5040& 1.3417& &&1.7309& 1.8097& 1.6846 \\
				& &  0.6487& 0.6660& 0.5963&   &&0.7706& 0.7997&  0.6954   \\
				\hline\\
				
		\end{tabular}}
		\end{center}
	\end{table}
	
	\begin{table}[!htbp]
		\renewcommand\thetable{4}
		\scriptsize \caption{\label{Table 4}Coverage probabilities for the
			pivotal quantities under MLEs  for $T=0.5$. The true values of $\alpha$ and $\lambda$ are $1.5$ and $0.75$, respectively.  }
		\begin{center}
			\scalebox{0.9}{
		\begin{tabular}{cccccccccccccccccc}
			\hline\\
			& &  & &90$\%$ confidence interval& & & &&95$\%$ confidence interval \\
			\hline\\
			(n,m)& scheme& &$Q_{1}$ & $Q_{2}$& $Q_{3}$& & & $Q_{1}$ & $Q_{2}$& $Q_{3}$\\
			\hline\\
			(35,10)& (0*9,25)& &0.738 & 0.759& 0.695& & & 0.805& 0.813& 0.772\\
			&  (0*5,5*5)& & 0.712& 0.714 & 0.665 & & &0.781 & 0.788 & 0.751\\
			&  (25,0*9)& & 0.622& 0.656& 0.601& & & 0.677& 0.733& 0.696 \\
			\hline\\
			(35,25)& (0*24,10)& & 0.772& 0.799& 0.756& & & 0.818& 0.853& 0.830\\
			& (0*20,2*5)& & 0.775& 0.815& 0.772& & & 0.834& 0.863& 0.833\\
			& (10,0*24)&  & 0.704& 0.738&  0.729& & & 0.757& 0.821& 0.804  \\
			\hline\\
			(40,10)& (0*9,30)&  & 0.738&  0.743& 0.646& & & 0.797&  0.796&  0.738  \\
			& (0*5,6*5)& & 0.735& 0.736& 0.685& & & 0.799& 0.799& 0.776\\
			& (30,0*9)& & 0.676& 0.740& 0.776 & & & 0.735& 0.802& 0.804\\
			\hline\\
			(40,30)& (0*29,10)& & 0.765& 0.807& 0.754& & & 0.825& 0.862& 0.825 \\
			& (0*20,1*10)& & 0.743& 0.765&  0.753& & & 0.799& 0.822& 0.832 \\
			& (10,0*29)& & 0.717& 0.754& 0.727& & & 0.780& 0.815&  0.810\\
			\hline\\
			
		\end{tabular}}
	\end{center}
	\end{table}

	\begin{table}[!htbp]
		\renewcommand\thetable{5}
		\scriptsize \caption{\label{Table 5}Average and MSE values of the
			estimates of $\alpha$ (true value=1.5) for $T=0.65$.  The first (second) rows corresponding to `U' and `B' are for the average (MSE) values of the estimates. }
		\begin{center}
		\scalebox{0.9}{
			\begin{tabular}{cccccccccccccccccc}
				\hline\\
				(n,m)& scheme &prior &$\widehat{\alpha}$&  &$\widehat{\alpha}_{LI}$&  & &$\widehat{\alpha}_{GE}$    &  && $\widehat{\alpha}_{SQ}$       \\
				\hline\\
				& & & &$p=-0.05$ & $p=0.5$ & $p=1$ &  $q=-0.5$ & $q=-0.25$ & $q=0.25$ && \\
				\hline\\
				(35,10)& (0*9,25)&U& 1.7557& 1.5770& 1.5330& 1.5026&  1.5492& 1.5386& 1.5199& & 1.5726  \\
				& & &0.0654&  0.0059&   0.0010& 0.0001&  0.0024& 0.0014& 0.0004& & 0.0053  \\
				& & B& & 1.6757&1.6131 &1.5655 & 1.6366&1.6210 &1.5920 & & 1.6697 \\
				& & & &  0.0308& 0.0128& 0.0043& 0.0186& 0.0146& 0.0085& & 0.0287    \\ [0.1 cm]
				& (0*5,5*5)&U&  1.7529&  1.5243&  1.4763& 1.4456&  1.4937& 1.4826&  1.4635& & 1.5193\\
				& & &0.0640&  0.0005&  0.0005&  0.0029&  0.0000& 0.0003& 0.0013& &  0.0004 \\
				& & B& & 1.5626&1.4838 &1.4327 & 1.5123&1.4937 &1.4616 & & 1.5545 \\
				& & & & 0.0039& 0.0002& 0.0045& 0.0002& 0.0000& 0.0015& & 0.0029 \\ [0.1 cm]
				& (25,0*9)&U& 1.6282& 1.5016& 1.4637& 1.4354&  1.4763& 1.4664& 1.4483& & 1.4979  \\
				& & & 0.0164& 0.0000&  0.0013&  0.0041&   0.0006&  0.0011& 0.0026& & 0.0000 \\
				& &B& & 1.5853&1.5436 &1.5095 &  1.5579&1.5465 &1.5249 & & 1.5874 \\
				& & & & 0.0072& 0.0019& 0.0001& 0.0033& 0.0021& 0.0006& & 0.0076    \\
				\hline\\
				(35,25)& (0*24,10)&U& 1.6164& 1.4812& 1.4407&  1.4108&  1.4540& 1.4433& 1.4241& & 1.4772 \\
				& & & 0.0135&  0.0003& 0.0035&  0.0079&  0.0021& 0.0032& 0.0057& & 0.0515   \\
				& & B& & 1.5715&1.5274 &1.4915 &  1.5422&1.5302 &1.5073 & & 1.5673 \\
				& & & &   0.0051& 0.0007& 0.0001& 0.0017& 0.0009& 0.0001& & 0.0045      \\ [0.1 cm]
				& (0*20,2*5)&U& 1.6367& 1.4982& 1.4552& 1.4238&   1.4695& 1.4584& 1.4383& & 1.4939  \\
				& & & 0.0187& 0.0000& 1.4552& 0.0058&  0.0009& 0.0017& 0.0038& & 0.0000  \\
				& & B& & 1.5769&1.5170 &1.4701 & 1.5374&1.5214 &1.4914 & & 1.5712 \\
				& & & &   0.0059& 0.0003& 0.0008& 0.0014& 0.0004& 0.0001& & 0.0051  \\ [0.1 cm]
				& (10,0*24)&U& 1.6291& 1.5042& 1.4621& 1.4308&   1.4761&  1.4651& 1.4450& & 1.5001  \\
				& &  & 0.0166& 0.0000&  0.0014& 0.0047&  0.0005& 0.0012&  0.0030& &  0.0000 \\
				& & B& & 1.5956&1.5515 &1.5152 &  1.5666&1.5546 &1.5316 & & 1.5915      \\
				& & & &  0.0091& 0.0026& 0.0002& 0.0044& 0.0029& 0.0010& & 0.0083     \\
				\hline\\
				(40,10)& (0*9,30)&U& 1.7836& 1.5228&  1.4745& 1.4453&  1.4923& 1.4815& 1.4635& & 1.5177  \\
				& & & 0.0804&  0.0005& 0.0006& 0.0029&  0.0000& 0.0003& 0.0013& & 0.0003 \\
				& & B& & 1.5922&1.4885 &1.4233 &  1.5266&1.5024 &1.4613 & & 1.5815 \\
				& & & & 0.0085& 0.0001& 0.0058& 0.0007& 0.0000& 0.0015& & 0.0066\\ [0.1 cm]
				& (0*5,6*5)&U&  1.7411& 1.5827&  1.5437& 1.5159&   1.5580& 1.5485& 1.5314& &  1.5788  \\
				& & & 0.0581& 0.0068& 0.0019& 0.0002&   0.0033& 0.0023& 0.0009& &  0.0062 \\
				& & B& & 1.6310&1.5781 &1.5387 &  1.5976&1.5844 &1.5603 & & 1.6258 \\
				& & & & 0.0171&0.0061 &0.0015 & 0.0095& 0.0071& 0.0036& & 0.0158 \\ [0.1 cm]
				& (30,0*9)&U& 1.6023& 1.5512& 1.5105&  1.4775&  1.5240& 1.5128& 1.4916& & 1.5474   \\
				& & & 0.0104& 0.0026& 0.0001&  0.0005&   0.0005& 0.0002& 0.0001& & 0.0022 \\
				& & B& & 1.5619&1.5176 &1.4815 &  1.5324&1.5201 &1.4969 & & 1.5577   \\
				& & & &  0.0038& 0.0003& 0.0003& 0.0011& 0.0004& 0.0000& & 0.0033  \\
				\hline\\
				(40,30)& (0*29,10)&U& 1.6069& 1.5067& 1.4731& 1.4470&  1.4841&  1.4751&  1.4583& & 1.5034  \\
				& & & 0.0114&  0.0001& 0.0007& 0.0027&   0.0002& 0.0006& 0.0017& & 0.0000 \\
				& & B& & 1.5186&1.4638 &1.4221 &  1.4818&1.4670 &1.4399 & & 1.5133 \\
				& & & &  0.0003& 0.0013& 0.0061& 0.0003& 0.0011& 0.0036& & 0.0002    \\ [0.1 cm]
				& (0*20,1*10)&U& 1.6031& 1.5033& 1.4703& 1.4448&  1.4811& 1.4723& 1.4558& & 1.5001  \\
				& & & 0.0106&  0.0000& 0.0008& 0.0030&  0.0003&  0.0007& 0.0019& & 0.0000 \\
				& & B& & 1.5255&1.4783 &1.4415 &  1.4938&1.4810 &1.4572 & & 1.5210      \\
				& & & &  0.0006& 0.0004& 0.0034& 0.0000& 0.0003& 0.0018& & 0.0004    \\ [0.1 cm]
				& (10,0*29)&U& 1.6292& 1.5227& 1.4846&  1.4556&   1.4974& 1.4873&  1.4687& & 1.5190 \\
				& & & 0.0167& 0.0005& 0.0002& 0.0019&  0.0000& 0.0002& 0.0009& & 0.0004 \\
				& & B& &  1.5850&1.5429 &1.5086 &  1.5573&1.5459 &1.5241 & & 1.5811 \\
				& & & &   0.0073& 0.0018& 0.0001& 0.0032& 0.0021& 0.0005& & 0.0065     \\
				\hline
		\end{tabular}}
	\end{center}
	\end{table}

	\begin{table}[!htbp]
		\renewcommand\thetable{6}
		\scriptsize \caption{\label{Table 6}Average and MSE values of the
			estimates of $\lambda$ (true value=0.75) for $T=0.65$.  The first (second) rows corresponding to `U' and `B' are for the average (MSE) values of the estimates. }
			\begin{center}
		\scalebox{0.9}{
			\begin{tabular}{cccccccccccccccccc}
				\hline\\
				(n,m)& scheme &prior &$\widehat{\lambda}$&  &$\widehat{\lambda}_{LI}$&   &   &$\widehat{\lambda}_{GE}$     & && $\widehat{\lambda}_{SQ}$       \\
				\hline\\
				& & & &$p=-0.05$ & $p=0.5$ & $p=1$ &  $q=-0.5$ & $q=-0.25$ & $q=0.25$ && \\
				\hline\\
				(35,10)& (0*9,25)&U& 0.8606& 0.7717& 0.7642& 0.7580&  0.7632& 0.7597& 0.7534& & 0.7710 \\
				& &   &0.0122& 0.0005&  0.0002& 0.0001&  0.0002& 0.0001& 0.0000& & 0.0004 \\
				& & B& & 0.7586&0.7508 &0.7445 & 0.7499&0.7463 &0.7401 & & 0.7579 \\
				& & & &  0.0001& 0.0000& 0.0002& 0.0000& 0.0000& 0.0001& & 0.0001    \\ [0.1 cm]
				& (0*5,5*5)&U&  0.8619&  0.7448& 0.7376& 0.7320&  0.7369& 0.7338& 0.7284& & 0.7441  \\
				& &&   0.0125& 0.0000& 0.0001& 0.0003&  0.0002& 0.0002&  0.0005& & 0.0000  \\
				& & B& & 0.7069&0.7015 &0.6976 & 0.7013&0.6993 &0.6962 & & 0.7063 \\
				& & & & 0.0018& 0.0023& 0.0027& 0.0024& 0.0026& 0.0028& & 0.0019   \\ [0.1 cm]
				& (25,0*9)&U&   0.8252& 0.7632& 0.7534& 0.7453&  0.7517& 0.7468& 0.7378& &  0.7623 \\
				& & & 0.0056& 0.0002& 0.0000& 0.0000&  0.0000&  0.0000&  0.0001& & 0.0002   \\
				& & B& & 0.7187&0.7102 &0.7034 &  0.7089&0.7049 &0.6981 & & 0.7179 \\
				& & & & 0.0010& 0.0015& 0.0021& 0.0016& 0.0020& 0.0026& & 0.0010    \\
				\hline\\
				(35,25)& (0*24,10)&U& 0.7977&  0.7750&  0.7690& 0.7637&  0.7677& 0.7644& 0.7580& & 0.7745  \\
				& & & 0.0022&  0.0006& 0.0003&  0.0001&  0.0003& 0.0002& 0.0001& &  0.0006  \\
				& & B& & 0.7355&0.7301 &0.7257 &  0.7290&0.7262 &0.7212 & & 0.7350 \\
				& & & & 0.0002& 0.0003& 0.0006& 0.0004& 0.0005& 0.0008& & 0.0002  \\ [0.1 cm]
				& (0*20,2*5)&U& 0.7942& 0.7556& 0.7498& 0.7449&  0.7486& 0.7455& 0.7396& & 0.7550 \\
				& & & 0.0019& 0.0000& 0.0000& 0.0000&   0.0000& 0.0000& 0.0001& & 0.0000 \\
				& & B& & 0.7219&0.7163 &0.7116 &  0.7151 &0.7123 &0.7071 & & 0.7213 \\
				& & & &  0.0008& 0.0011& 0.0015& 0.0012& 0.0014& 0.0018& & 0.0007  \\ [0.1 cm]
				& (10,0*24)&U& 0.8225&  0.7580& 0.7507& 0.7446&  0.7494& 0.7457& 0.7390& & 0.7573   \\
				& & & 0.0052& 0.0001&  0.0000& 0.0000&  0.0000& 0.0000& 0.0001& & 0.0001 \\
				& & B& & 0.7487&0.7424 &0.7371 &  0.7413&0.7381 &0.7324 & & 0.7481    \\
				& & & &   0.0000& 0.0001& 0.0002& 0.0001& 0.0001& 0.0003& & 0.0000   \\
				\hline\\
				(40,10)& (0*9,30)&U& 0.8880& 0.7580& 0.7518& 0.7471&  0.7514& 0.7489& 0.7447& & 0.7574  \\
				& & &  0.0190& 0.0001& 0.0000& 0.0000&  0.0000& 0.0000& 0.0000& & 0.0001  \\
				& & B& & 0.7149 &0.7075 &0.7024 &  0.7074&0.7048 &0.7010 & & 0.7142 \\
				& & & & 0.0012& 0.0018& 0.0022& 0.0018& 0.0020& 0.0023& & 0.0012   \\ [0.1 cm]
				& (0*5,6*5)&U& 0.8656&  0.7796&  0.7722& 0.7662&  0.7713& 0.7679& 0.7618& & 0.7789   \\
				& & & 0.0133& 0.0008&  0.0005&  0.0003&  0.0004& 0.0003& 0.0001& &  0.0008  \\
				& & B& & 0.7420&0.7348 &0.7294 &  0.7342&0.7312 &0.7260 & & 0.7413 \\
				& & & &  0.0001& 0.0002& 0.0004& 0.0002& 0.0003& 0.0006& & 0.0001       \\ [0.1 cm]
				& (30,0*9)&U& 0.8424& 0.7757& 0.7596& 0.7463&  0.7571& 0.7493& 0.7353& & 0.7741  \\
				& & & 0.0085&  0.0006& 0.0001& 0.0000&  0.0001&  0.0000& 0.0002& & 0.0006  \\
				& & B& & 0.7031&0.6938 &0.6868 &  0.6928&0.6889 &0.6826 & & 0.7022 \\
				& & & &  0.0021& 0.0031& 0.0039& 0.0032& 0.0037& 0.0045& & 0.0022       \\
				\hline\\
				(40,30)& (0*29,10)&U& 0.7917& 0.7529& 0.7478& 0.7434&   0.7466&  0.7439& 0.7386& & 0.7524  \\
				& & & 0.0017& 0.0000& 0.0000& 0.0000&   0.0000& 0.0000&  0.0001& & 0.0000 \\
				& & B& & 0.7086&0.7031 &0.6986 &  0.7020&0.6993 &0.6944 & & 0.7081 \\
				& & & & 0.0017& 0.0021& 0.0026& 0.0022& 0.0025& 0.0030& & 0.0017    \\ [0.1 cm]
				& (0*20,1*10)&U& 0.7982& 0.7603& 0.7552& 0.7508&  0.7541& 0.7513& 0.7461& & 0.7599  \\
				& & & 0.0023& 0.0001& 0.0000& 0.0000&  0.0000& 0.0000& 0.0000& & 0.0001 \\
				& & B& & 0.7137&0.7083 &0.7038 &  0.7071&0.7044 &0.6995 & & 0.7131    \\
				& & & &  0.0013& 0.0017& 0.0021& 0.0018& 0.0020& 0.0025& & 0.0013       \\ [0.1 cm]
				& (10,0*29)&U& 0.7995& 0.7557& 0.7495& 0.7441&   0.7481& 0.7448& 0.7385& &  0.7552  \\
				& & & 0.0024&  0.0000& 0.0000& 0.0000&  0.0000& 0.0000& 0.0001& & 0.0000 \\
				& & B& & 0.7288&0.7229 &0.7180 &  0.7217&0.7187 &0.7132 & & 0.7282    \\
				& & & &  0.0004& 0.0007& 0.0010& 0.0008& 0.0009& 0.0013& & 0.0005     \\
				\hline\\
		\end{tabular}}
	\end{center}
	\end{table}

	\begin{table}[!htbp]
		\renewcommand\thetable{7}
		\scriptsize \caption{\label{Table 7}Average lengths of the interval
			estimates of $\alpha$ and $\lambda$ for $T=0.65$. The first (second) row corresponding to the scheme is for  $\alpha$ ($\lambda$).  The true values of $\alpha$ and $\lambda$ are $1.5$ and $0.75$, respectively.  }
		\begin{center}
		\scalebox{0.9}{
			\begin{tabular}{cccccccccccccccccc}
				\hline\\
				& &  & 90$\%$ confidence interval&  & & &&95$\%$ confidence interval \\
				\hline\\
				(n,m)& scheme& NA & NL&  HPD& & &NA & NL& HPD\\
				\hline\\
				(35,10)& (0*9,25)& 1.5116& 1.5587& 1.4545 & & & 1.7956& 1.8749& 1.7695\\
				& &  0.6340&  0.6485& 0.6191 & & & 0.7532& 0.7774& 0.7291 \\
				& (0*5,5*5)&  1.6574& 1.7198& 1.6101 & & & 1.9688& 2.0739& 1.9365  \\
				& & 0.6750& 0.6924&  0.6606& & & 0.8018& 0.8311&  0.7447    \\
				& (25,0*9)& 1.3392& 1.3773& 1.3093& & & 1.5908& 1.6548& 1.5479  \\
				& & 0.6647& 0.6828&  0.6471& & & 0.7896& 0.8201& 0.7351    \\
				\hline\\
				(35,25)& (0*24,10)& 1.3929& 1.4364& 1.2914& & & 1.6546& 1.7278& 1.6122  \\
				& &  0.4968& 0.5049& 0.4770& & & 0.5902& 0.6037& 0.5617 \\
				& (0*20,2*5)&  1.4346& 1.4810& 1.4120& & & 1.7042& 1.7822& 1.6506 \\
				& & 0.4972&  0.5054& 0.4689& & & 0.5906& 0.6043& 0.5531  \\
				& (10,0*24)& 1.4019& 1.4456& 1.3306& & & 1.6653& 1.7388& 1.6152   \\
				& &  0.5873& 0.5998& 0.5423& & & 0.6976& 0.7187&  0.6800 \\
				\hline\\
				(40,10)& (0*9,30)& 1.7256& 1.7936& 1.6322& & & 2.0498& 2.1644& 1.9857  \\
				& &  0.6725& 0.6887& 0.6329 & & & 0.7988& 0.8260& 0.7401   \\
				& (0*5,6*5)&  1.4019& 1.4401& 1.3476& & & 1.6653& 1.7295& 1.6023   \\
				& &  0.6230& 0.6365&  0.5875 & & & 0.7401& 0.7628& 0.7265  \\
				& (30,0*9)& 1.3029& 1.3391& 1.2550& & & 1.5477& 1.6086& 1.4705  \\
				& &  0.8422& 0.8777& 0.8047& & & 1.0004& 1.0603& 0.9395   \\
				\hline\\
				(40,30)& (0*29,10)& 1.2337& 1.2643&  1.2128& & & 1.4655& 1.5168& 1.4127    \\
				& &  0.4714& 0.4784&  0.4582& & & 0.5601& 0.5718& 0.5445  \\
				& (0*20,1*10)&  1.2210& 1.2508& 1.2088& & & 1.4505& 1.5004& 1.4260 \\
				& &  0.4710& 0.4779& 0.4519& & & 0.5595& 0.5711&  0.5225  \\
				& (10,0*29)& 1.3151& 1.3510& 1.2714& & & 1.5621& 1.6227& 1.5202   \\
				& & 0.5237& 0.5331& 0.5773& & & 0.6221& 0.6379& 0.5958 \\
				
				\hline
		\end{tabular}}
	\end{center}
	\end{table}
	
	\begin{table}[!htbp]
		\renewcommand\thetable{8}
		\scriptsize \caption{\label{Table 8}Coverage probabilities for the
			pivotal quantities under the MLEs for $T=0.65$. The true values of $\alpha$ and $\lambda$ are $1.5$ and $0.75$, respectively.  }
		\begin{center}
			\scalebox{0.9}{
		\begin{tabular}{cccccccccccccccccc}
			\hline\\
			& &  & &90$\%$ confidence interval& & & &&95$\%$ confidence interval \\
			\hline\\
			(n,m)& scheme& &$Q_{1}$ & $Q_{2}$& $Q_{3}$& & & $Q_{1}$ & $Q_{2}$& $Q_{3}$\\
			\hline\\
			(35,10)& (0*9,25)& &0.758& 0.758& 0.757& & & 0.821& 0.804& 0.836 \\
			& (0*5,5*5)& & 0.785& 0.780& 0.789& & & 0.845& 0.835& 0.859    \\
			& (25,0*9)& &   0.643& 0.675& 0.636& & & 0.714& 0.757& 0.712  \\
			\hline\\
			(35,25)& (0*24,10)& &  0.797& 0.822& 0.787& & &  0.853& 0.874& 0.841   \\
			& (0*20,2*5)& & 0.775& 0.801& 0.770& & & 0.834& 0.861& 0.835 \\
			& (10,0*24)& & 0.736& 0.765& 0.779& & & 0.792&  0.832& 0.834\\
			\hline\\
			(40,10)& (0*9,30)& & 0.817& 0.785& 0.779& & &  0.873& 0.833& 0.856\\
			& (0*5,6*5)& & 0.743& 0.728& 0.750& & & 0.801& 0.787& 0.826\\
			& (30,0*9)& & 0.633& 0.660& 0.748& & &  0.692& 0.743& 0.799\\
			\hline\\
			(40,30)& (0*29,10)& & 0.752& 0.783& 0.767& & & 0.818& 0.842& 0.838 \\
			& (0*20,1*10)& & 0.753& 0.781& 0.772& & & 0.823& 0.852& 0.839\\
			& (10,0*29)& & 0.734& 0.767& 0.772& & & 0.811& 0.832& 0.838\\
			\hline
			
		\end{tabular}}
	\end{center}
	\end{table}

	\section{Real data analysis}
	In this section, a real data set of breaking stress of carbon fibres given by \cite{nichols2006bootstrap} is considered and analysed. This data set describes $100$ observations on breaking stress of carbon fibers (in GPa.). The data set is given below.\\
	\\
	 ---------------------------------------------------------------------------------------------------------------------\\
	3.7, 2.74, 2.73, 2.5, 3.6, 3.11, 3.27, 2.87, 1.47, 3.11, 4.42, 2.41, 3.19, 3.22, 1.69, 3.28, 3.09, 1.87, 3.15, 4.9, 3.75, 2.43, 2.95, 2.97, 3.39, 2.96, 2.53, 2.67, 2.93, 3.22, 3.39, 2.81, 4.2, 3.33, 2.55, 3.31, 3.31, 2.85, 2.56, 3.56, 3.15, 2.35, 2.55, 2.59, 2.38, 2.81, 2.77, 2.17, 2.83, 1.92, 1.41, 3.68, 2.97, 1.36, 0.98, 2.76, 4.91, 3.68, 1.84, 1.59, 3.19, 1.57, 0.81, 5.56, 1.73, 1.59, 2, 1.22, 1.12, 1.71, 2.17, 1.17, 5.08, 2.48, 1.18, 3.51, 2.17, 1.69, 1.25, 4.38, 1.84, 0.39, 3.68, 2.48, 0.85, 1.61, 2.79, 4.7, 2.03, 1.8, 1.57, 1.08, 2.03, 1.61, 2.12, 1.89, 2.88, 2.82, 2.05, 3.65\\
	 ----------------------------------------------------------------------------------------------------------------------
	\\
	To study the goodness of fit test of this data for the LE distribution, negative log-likelihood
	criterion, Alkaikes-information criterion (AIC), AICc, Bayesian
	information criterion (BIC), Kolmogorov–Smirnov (K–S) distance and corresponding $p$-values are computed. The values of these test
	statistics are presented in Table $9$, which suggests that the
	LE distribution provides the best fit to the data
	comparison to the exponential distribution (ED), inverse exponential distribution (IED), exponentiated exponential distribution (EED), inverse
	Weibull distribution (IWD) and gamma distribution (GD).

	 For the purpose of goodness-of-fit test, we consider
	different plots in Figures $3$, $4$, $5$ and $6$  to show that the LE distribution fits the real data set well.
	In Figure $3$, $Q$-$Q$ plots are used to compare the given data set to the theoretical model. Here, $Q$-$Q$ plot represents the points
	$(F^{-1}(i/(n+1);\hat{\alpha},\hat{\lambda}),x_{(i)})$, where $i=1,\ldots,n$, and $x_{(i)}$
	represents the ordered values of given real data. In Figure $4$ , $P$-$P$
	plot represents the points
	$(F(x_{i};\hat{\alpha},\hat{\lambda}),F_{n}(x_{(i)}))$, where $F_{n}(x_{(i)})= (1/n)\sum_{i=1}^{n}I(X \leq
	x_{(i)})$ is the empirical distribution function and I is the indicator function. In Figure $5$, empirical cumulative distribution function (ECDF) plots have been made
	which represent the CDFs of the corresponding distributions for observed data and it is compared with theoretical CDFs. In Figure $6$, plots of histogram of real data set and theoretical densities of associated distributions have been made. From all these it has been noticed that LE distributions fits the real data set better than the other above mentioned distributions.

	In Table $10$, the MLEs and  the Bayes estimates are obtained for the parameters of the LE distributions based on the real data set. In Table $11$, $90\%$ and $95\%$ approximate confidence and HPD credible intervals are tabulated. From Table $10$ , it is clear that the values of the estimates remain moderately close to each other. From Table $11$, it is clear that the length of the HPD credible intervals are smaller than that of the asymptotic confidence intervals.  In a lifetime experiment, the choice of an optimal censoring scheme from a class of all possible censoring schemes has gained considerable attention in recent years. For example, one can see \cite{pradhan2013inference} and \cite{bhattacharya2017computation}. $A$-optimality and $D$-optimality criteria are mostly used criteria to choose an optimal censoring scheme for the distributions having more than one parameter. In the case of the $A$-optimality criterion, we have to minimize the trace, and for $D$-optimality, we have to minimize the determinant of the inverse of the observed Fisher information matrix. Comparison of different censoring schemes based on these two criteria is reported in Table $10$. Here, $A$-optimality is considered as C-I, and $D$-optimality is considered as C-II. Based on these two criterion, when $(n,m)$ is considered as $(100,20)$ then $(0*19,80*1)$ is optimal among other selected censoring schemes and when $(n,m)$ is considered as $(100,40)$ then $(0*39,60*1)$ is optimal among other selected censoring schemes, which have been reported in Table $10.$ 
	
	\begin{table}[!htbp]
		\renewcommand\thetable{9} \label{Table 9}
		\scriptsize \caption{The MLEs and statistics for
			goodness-of-fit test.}
		\begin{center}
			\scalebox{0.8}{
			\begin{tabular}{cccccccccccccccccc}
				\hline\\
				&Model  & $\widehat\alpha$ &  $\widehat\lambda$&   &-logL &&  &AIC     & &AICC   &&BIC && K-S distance && p-value \\
				\hline\\
				&LED &  3.0172 & 0.2750 & &  143.2473& &  &290.4946 & & 290.6183& & 295.7049 && 0.0837&& 0.4601\\
				\\
				& ED & --------   &  0.3185 & & 196.3709& & & 394.7418& & 394.7826& & 397.3469 && 0.2630 && 1.35 $\times 10^{-6}$\\
				\\
				& IED&  -------- &  2.1399 & & 199.3956 & & & 400.7912& & 400.8320& & 403.3963 && 0.3577 && 5.83 $\times 10^{-12}$\\
				\\
				& EED&  7.7854 & 1.0131 & & 146.1823 & & & 296.3646& &296.4883 & & 301.5749&& 0.1078 && 0.1819\\
				\\
				& IWD & 1.7689 & 3.0884 & & 173.1440& & & 350.2880& & 350.4117& & 355.4983 && 0.1777 && 0.0031\\
				\\
				& GD & 5.9529 & 2.2710 & & 168.9527 & & &341.9504 & & 342.0291& &347.1157 && 0.0935 && 0.3254\\
				\hline
		\end{tabular}}
	\end{center}
	\end{table}
	
	\begin{table}[!htbp]
		\renewcommand\thetable{10} 
		\scriptsize \caption{\label{Table 10}Values of estimates of $\alpha$
			and $\lambda$  for the real data set. }
		\scalebox{0.8}{
			\begin{tabular}{cccccccccccccccccc}
				\hline\\
				(n,m)& $T$ & scheme&  Prior& $\theta$ &MLE&  &LLF&  & &GELF     & & SELF & C-I& C-II      \\
				\hline
				& & & & & &$p=-0.05$ & $p=0.5$ & $p=1$  & $q=-0.5$ & $q=-0.25$ & $q=0.25$ & & &\\
				\hline\\
				(100,20)& 2& (0*19,80*1)& U&  $\alpha$& 2.8522& 2.9576& 2.9574& 2.9573&  2.9581& 2.9579& 2.9577&  2.9576 & 0.3222& $1.24 \times 10^{-4}$  \\
				& & & &  $\lambda$ & 0.2833& 0.2725& 0.2724& 0.2722&  0.2726& 0.2725& 0.2724&  0.2725 &\\
				& & & B&  $\alpha$& & 2.9340& 2.9338& 2.9335&  2.9339& 2.9339& 2.9338&  2.9340  &  \\
				& & & &  $\lambda$ &  & 0.2875& 0.2874& 0.2873&  0.2871& 0.2869& 0.2865&  0.2875 & \\
				& & (0*10,8*10)& U& $\alpha$& 3.5384& 3.4594& 3.4570& 3.4549&  3.4586&  3.4583& 3.4576&  3.4592  & 0.4853& $1.58 \times 10^{-4}$ \\
				& & & & $\lambda$ & 0.3285& 0.2840& 0.2839&  0.2838& 0.2837& 0.2836&  0.2834 & 0.2840 & \\
				& & & B&  $\alpha$& & 3.4806& 3.4798& 3.4791& 3.4803& 3.4802& 3.4800&  3.4806  &  \\
				& & &  &$\lambda$ &  & 0.2927& 0.2925& 0.2923&  0.2920& 0.2917& 0.2911&  0.2926 &\\
				& & (80*1,0*19)& U& $\alpha$& 4.6082 & 4.1443& 4.1293& 4.1159&  4.1396&  4.1380& 4.1347&  4.1429 & 0.6534& $6.59 \times 10^{-4}$ \\
				& & & & $\lambda$ & 0.5377 & 0.4868& 0.4866& 0.4863& 0.2837& 0.4863&  0.4860& 0.4855 & \\
				& & & B&  $\alpha$& & 4.4203& 4.4161& 4.4124& 4.4191& 4.4186& 4.4177&  4.4199 & \\
				& & & & $\lambda$ &  & 0.4767& 0.4758& 0.4749&  0.4749& 0.4740& 0.4733&  0.4766 &\\
				\hline\\
				(100,40)& 2& (0*39,60*1)& U&  $\alpha$& 2.6357& 2.7402& 2.7391& 2.7381&  2.7398& 2.7396& 2.7392&  2.7401 & 0.1845& $0.55 \times 10^{-4}$   \\
				& & & &  $\lambda$ & 0.2699& 0.2831& 0.2829& 0.2828&  0.2826& 0.2824& 0.2820&  0.2830 &\\
				& & & B&  $\alpha$& & 2.7146& 2.7133& 2.7121&  2.7140& 2.7138& 2.7134&  2.7144 &   \\
				& & & & $\lambda$ &  & 0.2884& 0.2883& 0.2881&  0.2883& 0.2882& 0.2880&  0.2884 &\\
				& & (0*20,3*20)& U& $\alpha$& 2.8287& 2.8668& 2.8653& 2.8640&  2.8661& 2.8659& 2.8655&  2.8666 & 0.2113& $0.60 \times 10^{-4}$   \\
				& & & &  $\lambda$ & 0.2831& 0.2715& 0.2714& 0.2712&  0.2713& 0.2712& 0.2711&  0.2715 & \\
				& & & B&  $\alpha$& & 2.8372& 2.8369& 2.8366&  2.8371& 2.8370& 2.8370&  2.8372 &   \\
				& & & & $\lambda$ &  & 0.2621& 0.2620& 0.2619&  0.2618& 0.2616& 0.2613&  0.2621 & \\
				& & (60*1,0*39)& U& $\alpha$& 3.5248& 3.4427& 3.4406& 3.4391&  3.4416& 3.4414& 3.4409&  3.4421 & 0.2795& $1.50 \times 10^{-4}$ \\
				& & & &  $\lambda$ & 0.4144& 0.3921& 0.3919& 0.3918&  0.3917& 0.3916& 0.3913&  0.3920 & \\
				& & & B&  $\alpha$& & 3.4522& 3.4506& 3.4490&  3.4517& 3.4515& 3.4510&  3.4521  &  \\
				& & & & $\lambda$ &  & 0.4017& 0.4016& 0.4014&  0.4014& 0.4019& 0.4010&  0.4018 &\\
				\hline
		\end{tabular}}
	\end{table}
	
	\begin{table}[!htbp]
		 \begin{center}
		\renewcommand\thetable{11}
		\scriptsize \caption{Lengths of interval estimates of
			$\alpha$ and $\lambda$ for the real data set.  }
		\scalebox{0.9}{
			\begin{tabular}{cccccccccccccccccc}
				\hline
				& & & & &90$\%$ confidence interval&  & & & &95$\%$ confidence interval \\
				\hline\\
				(n,m)& T&  scheme& $\theta$& NA & NL&  HPD& & &NA & NL& HPD\\
				\hline\\
				(100,20)& 2& (0*19,80*1)& $\alpha$& 1.8704& 1.9041& 1.0766& & & 2.2218& 2.2784& 1.1828 \\
				& & & $\lambda$&  0.0994& 0.0999& 0.0896& & & 0.1181& 0.1189& 0.0962 \\
				& & (0*10,8*10)& $\alpha$& 2.2972& 2.3377& 1.2671& & & 2.7288& 2.7969& 1.2813 \\
				& & & $\lambda$&  0.0901& 0.0904& 0.0783& & & 0.1069& 0.1075& 0.0856 \\
				& & (80*1,0*19)& $\alpha$& 2.6654& 2.7027& 1.6717& & & 3.1662 & 3.2288& 1.7505 \\
				& & & $\lambda$&  0.1058& 0.1060& 0.0970& & & 0.1257& 0.1260& 0.1134 \\
				\hline\\
				(100,40)& 2& (0*39,60*1)& $\alpha$& 1.4159& 1.4330& 0.7204& & & 1.6819& 1.7106& 0.8520 \\
				& & & $\lambda$&  0.0721& 0.0724& 0.0677& & & 0.0857& 0.0861& 0.0798 \\
				& & (0*20,3*20)& $\alpha$& 1.5153& 1.5335& 0.7437& & & 1.8000& 1.8305& 0.7642 \\
				& & & $\lambda$&  0.0699& 0.0702& 0.0627& & & 0.0831& 0.0834& 0.0792 \\
				& & (60*1,0*39)& $\alpha$& 1.7432& 1.7611& 0.8626& & & 2.0707& 2.1006& 0.9248 \\
				& & & $\lambda$&  0.0767& 0.0769& 0.0660& & & 0.0911& 0.0912& 0.0865 \\
				\hline
		\end{tabular}}
	\end{center}
	\end{table}
	
	
	\begin{figure}[h!]
		\begin{center}
			\subfigure[]{\label{c1}\includegraphics[width=1.85in]{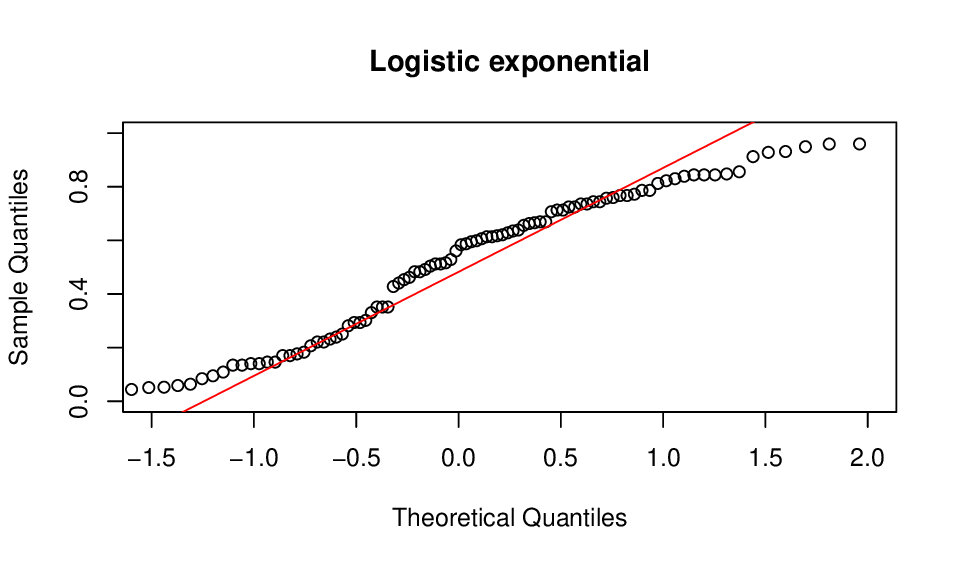}}
			\subfigure[]{\label{c1}\includegraphics[width=1.85in]{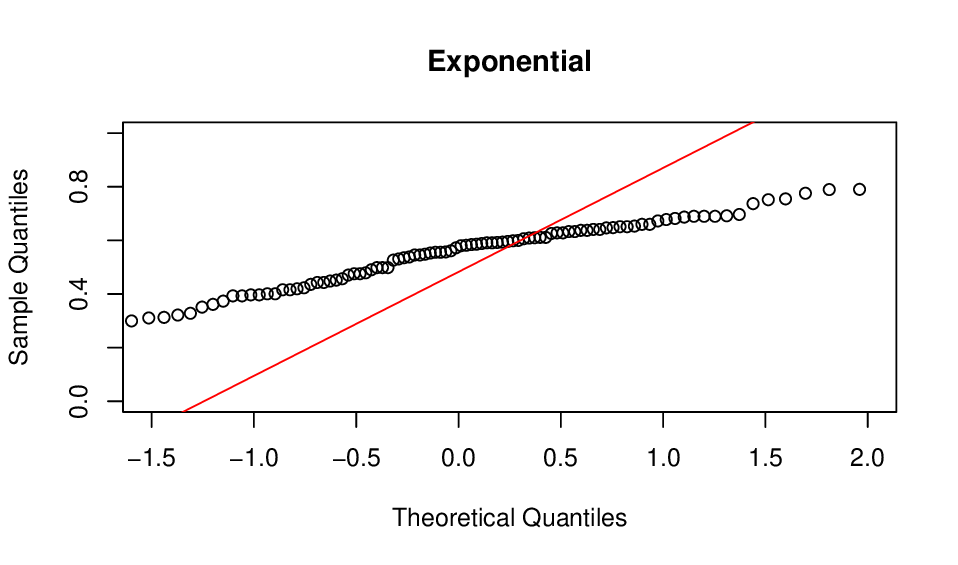}}
			\subfigure[]{\label{c1}\includegraphics[width=1.85in]{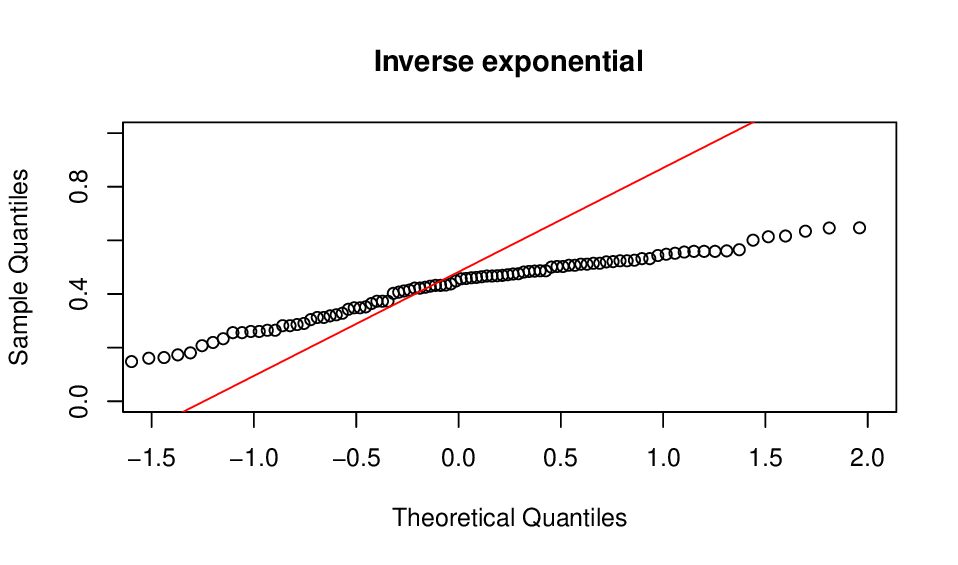}}
			\subfigure[]{\label{c1}\includegraphics[width=1.85in]{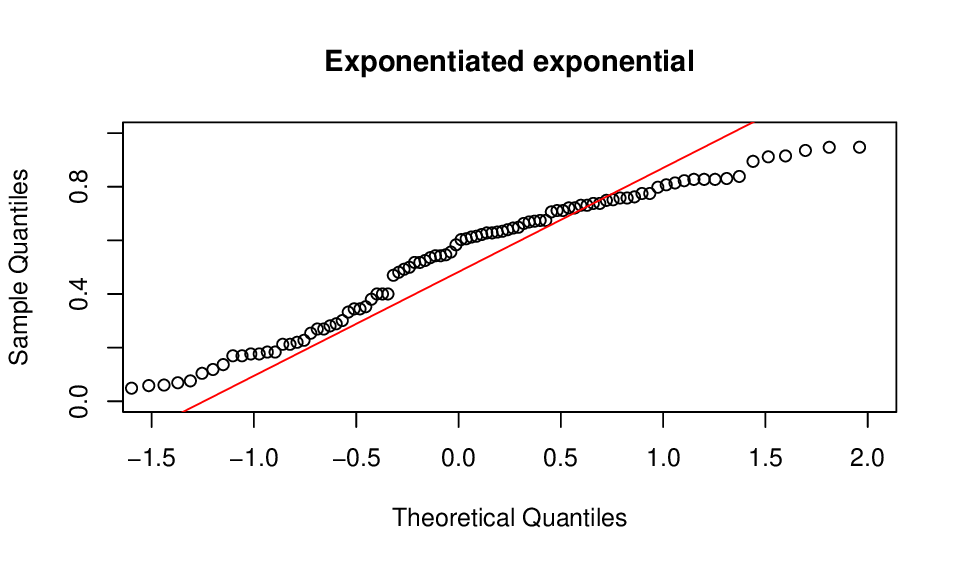}}
			\subfigure[]{\label{c1}\includegraphics[width=1.85in]{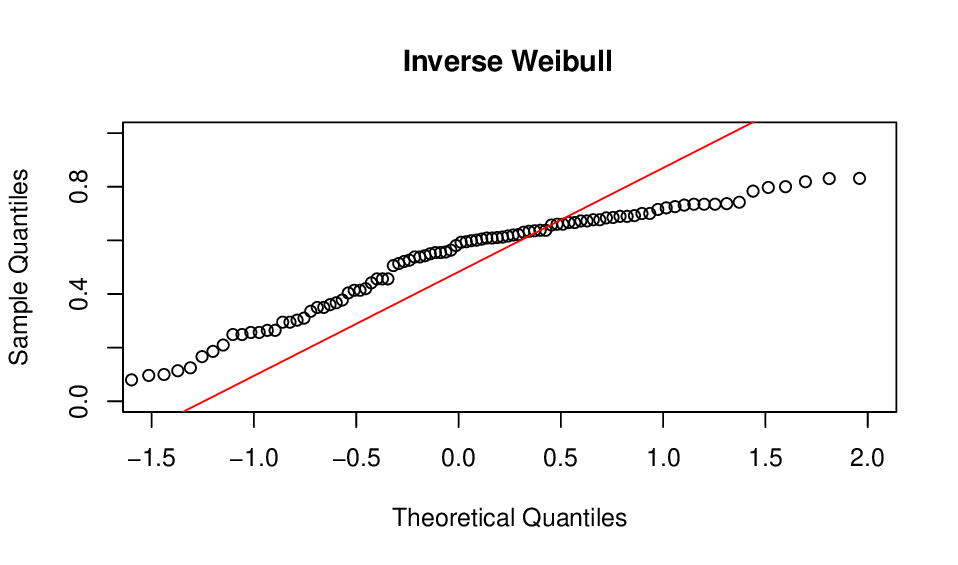}}
			\subfigure[]{\label{c1}\includegraphics[width=1.85in]{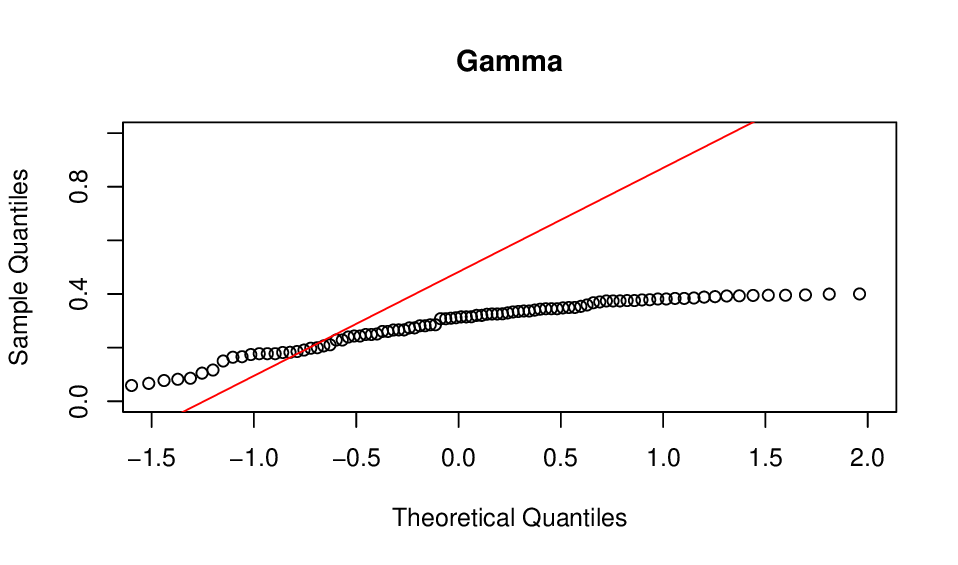}}
			\caption{Q-Q plots for various distributions fitted to the assumed real data set.}
		\end{center}
	\end{figure}
	\begin{figure}[h!]
	\begin{center}
		\subfigure[]{\label{c1}\includegraphics[width=1.85in]{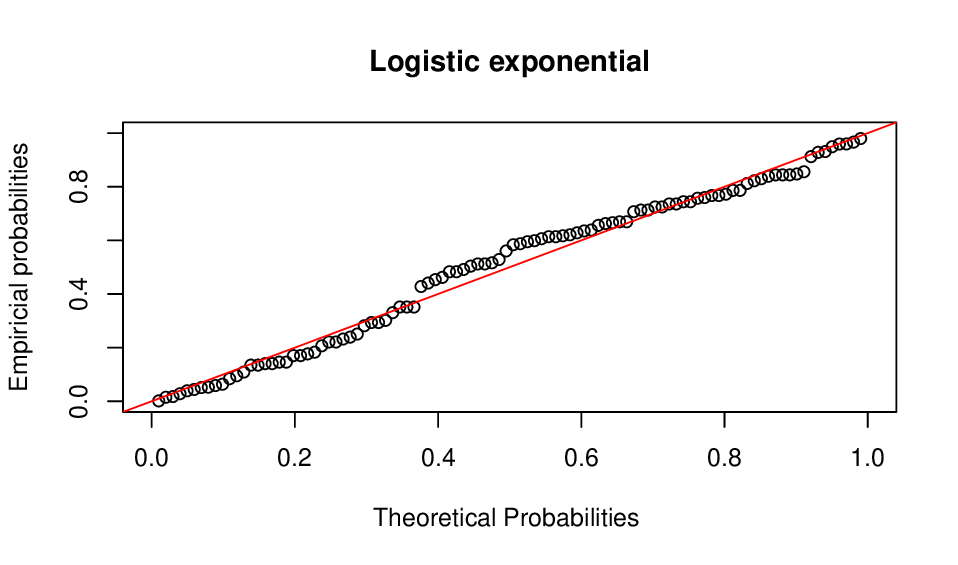}}
		\subfigure[]{\label{c1}\includegraphics[width=1.85in]{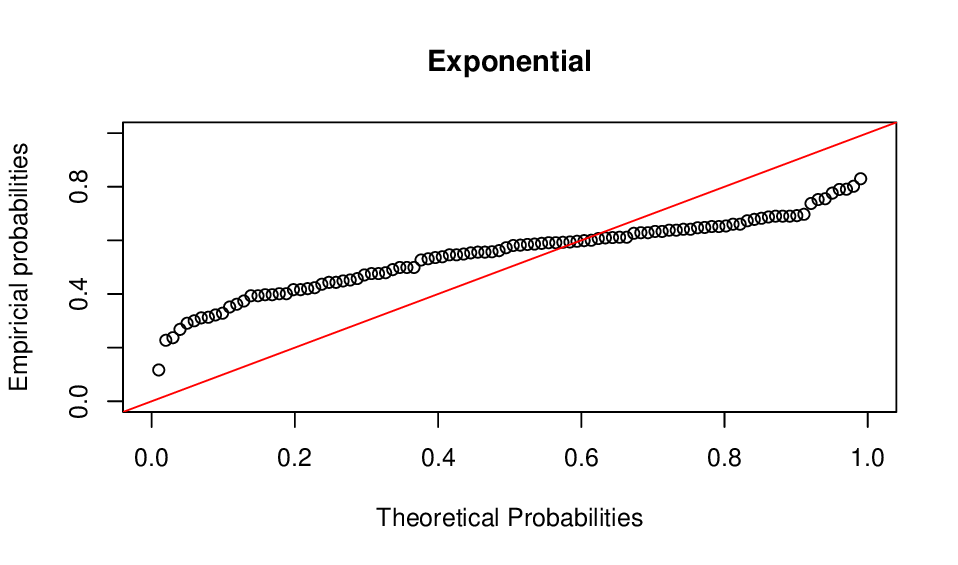}}
		\subfigure[]{\label{c1}\includegraphics[width=1.85in]{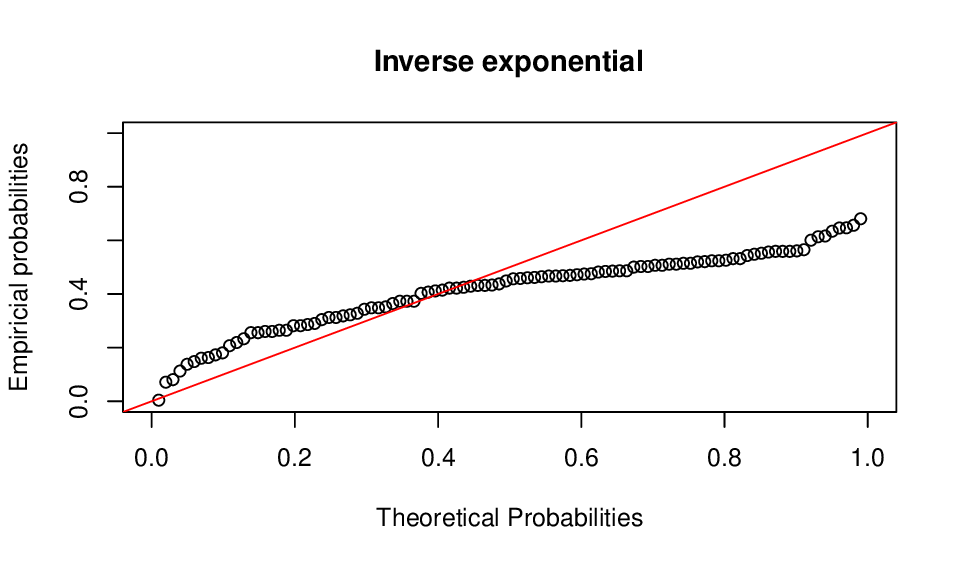}}
		\subfigure[]{\label{c1}\includegraphics[width=1.85in]{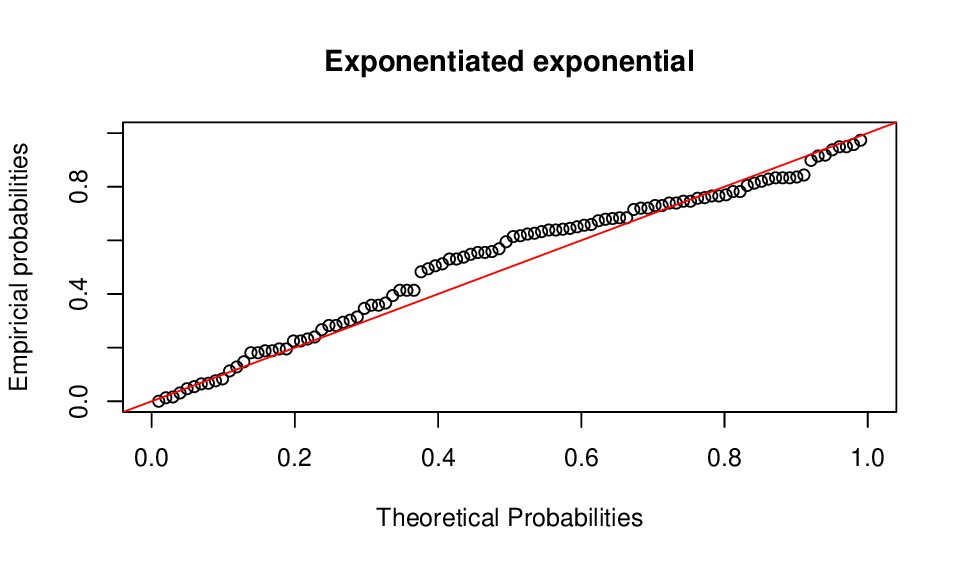}}
		\subfigure[]{\label{c1}\includegraphics[width=1.85in]{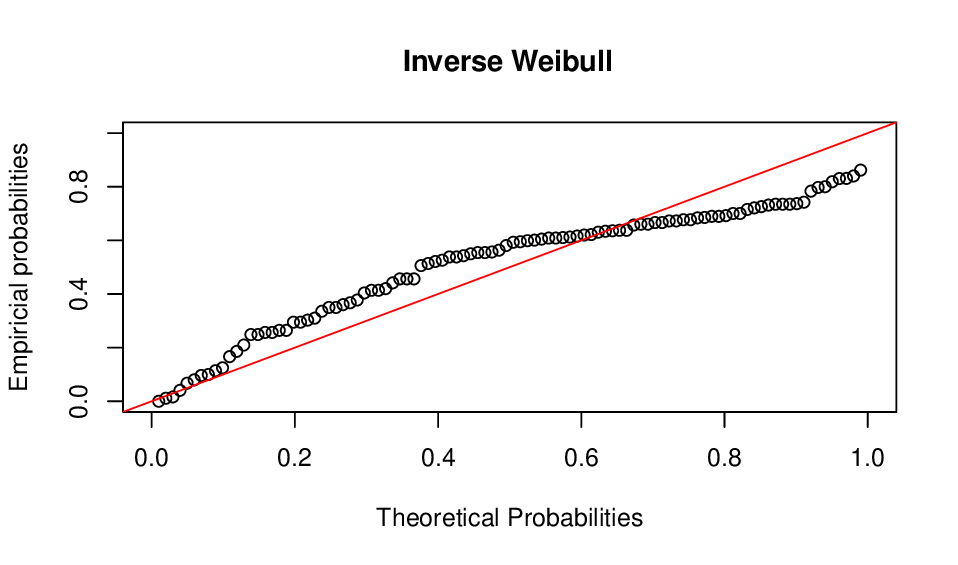}}
		\subfigure[]{\label{c1}\includegraphics[width=1.85in]{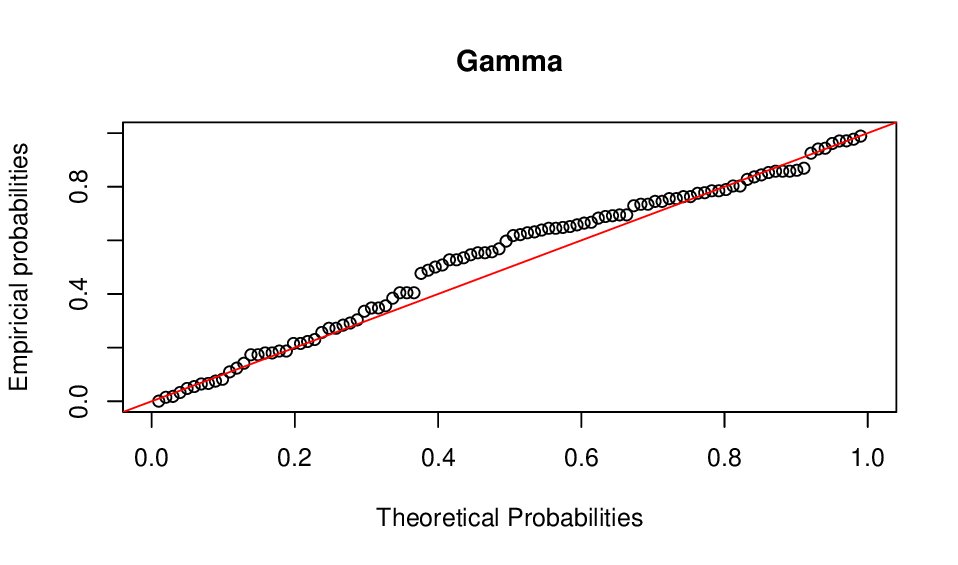}}
		\caption{The P-P plots for various distributions fitted to the assumed real data set.}
	\end{center}
\end{figure}

	\begin{figure}[h!]
	\begin{center}
		\subfigure[]{\label{c1}\includegraphics[width=1.85in]{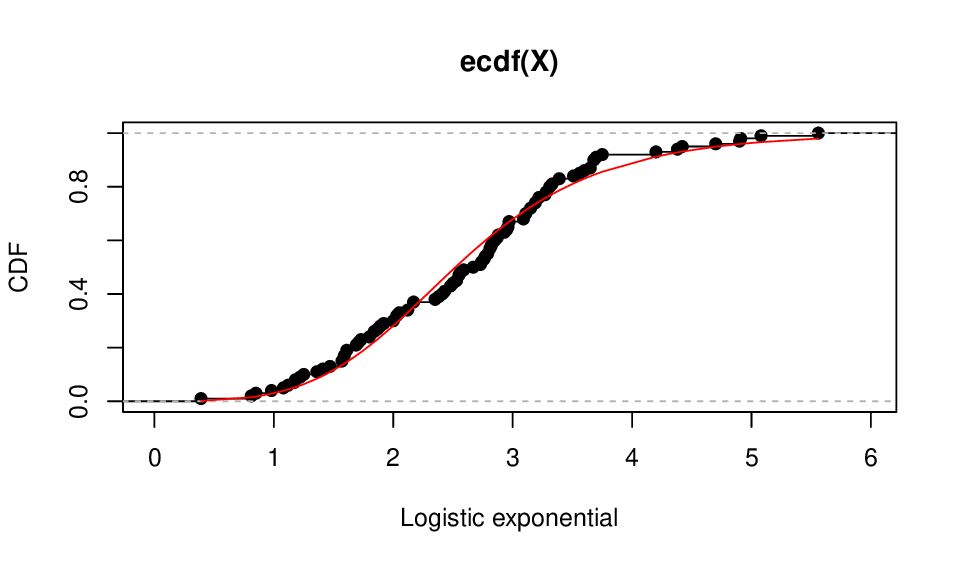}}
		\subfigure[]{\label{c1}\includegraphics[width=1.85in]{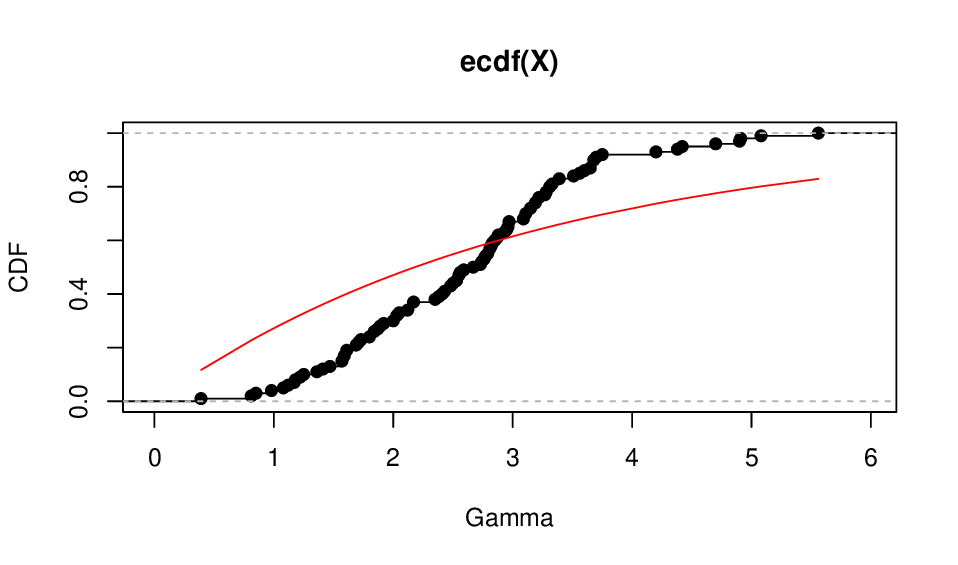}}
		\subfigure[]{\label{c1}\includegraphics[width=1.85in]{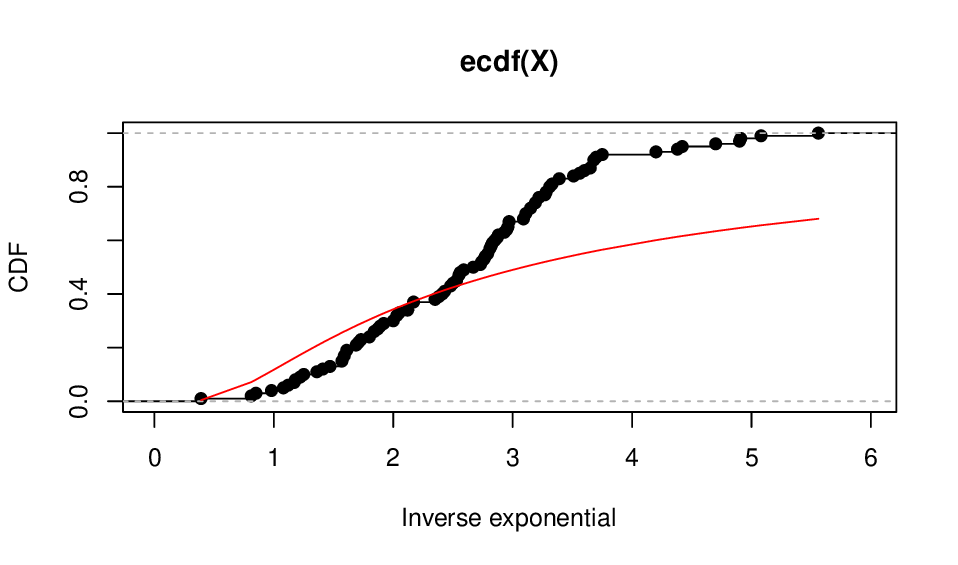}}
		\subfigure[]{\label{c1}\includegraphics[width=1.85in]{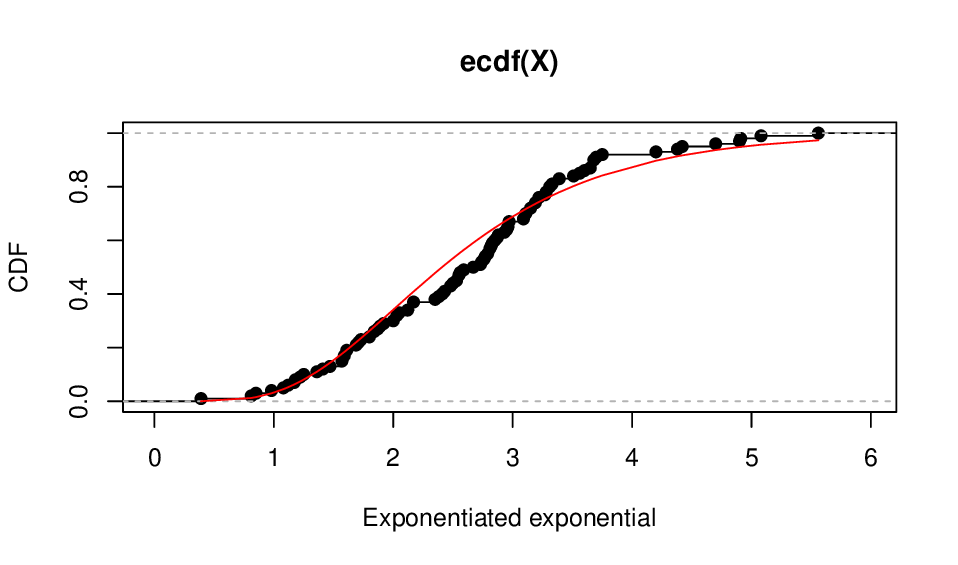}}
		\subfigure[]{\label{c1}\includegraphics[width=1.85in]{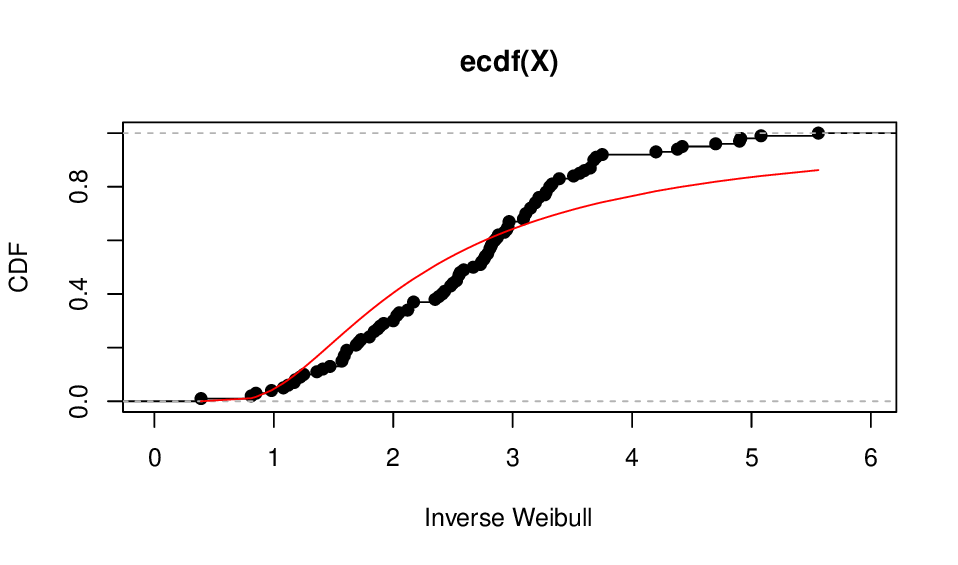}}
		\subfigure[]{\label{c1}\includegraphics[width=1.85in]{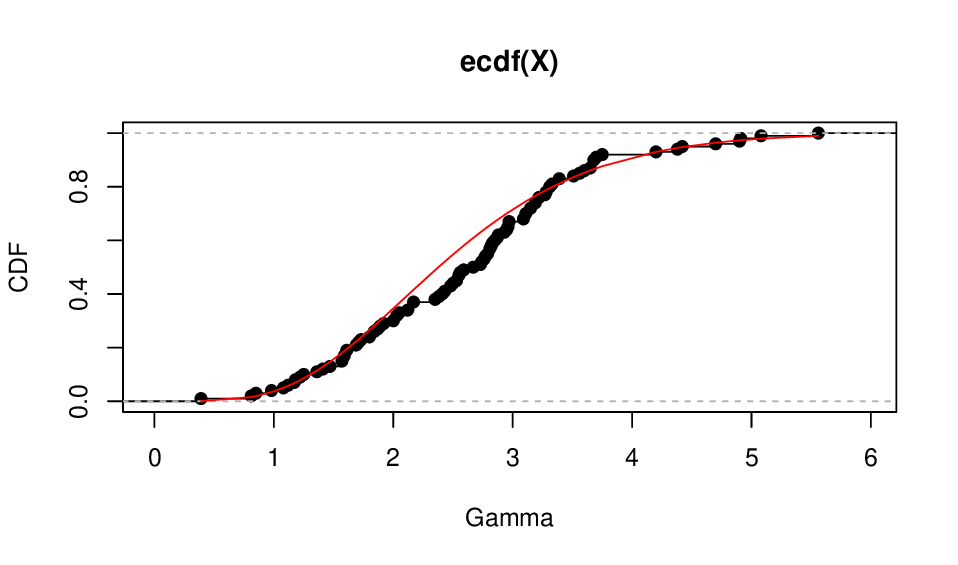}}
		\caption{ The ECDF vs. CDF plots for various distributions fitted to the assumed real data set.}
	\end{center}
\end{figure}

	\begin{figure}[h!]
	\begin{center}
		\subfigure[]{\label{c1}\includegraphics[width=1.85in]{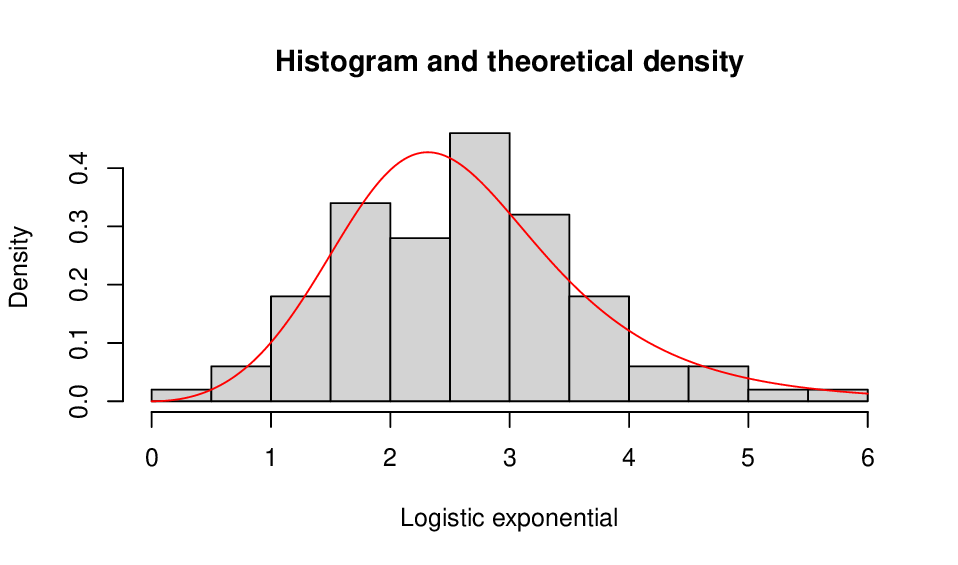}}
		\subfigure[]{\label{c1}\includegraphics[width=1.85in]{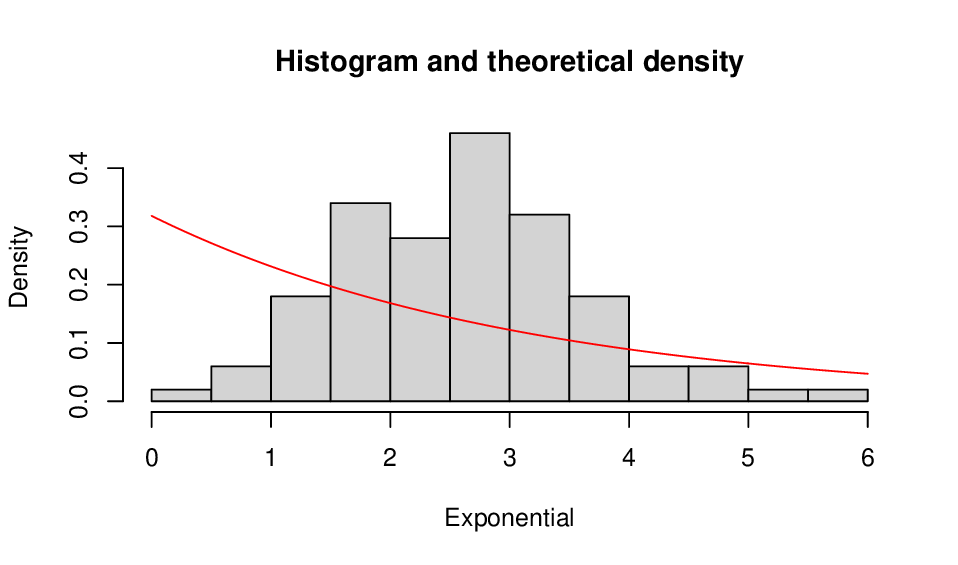}}
		\subfigure[]{\label{c1}\includegraphics[width=1.85in]{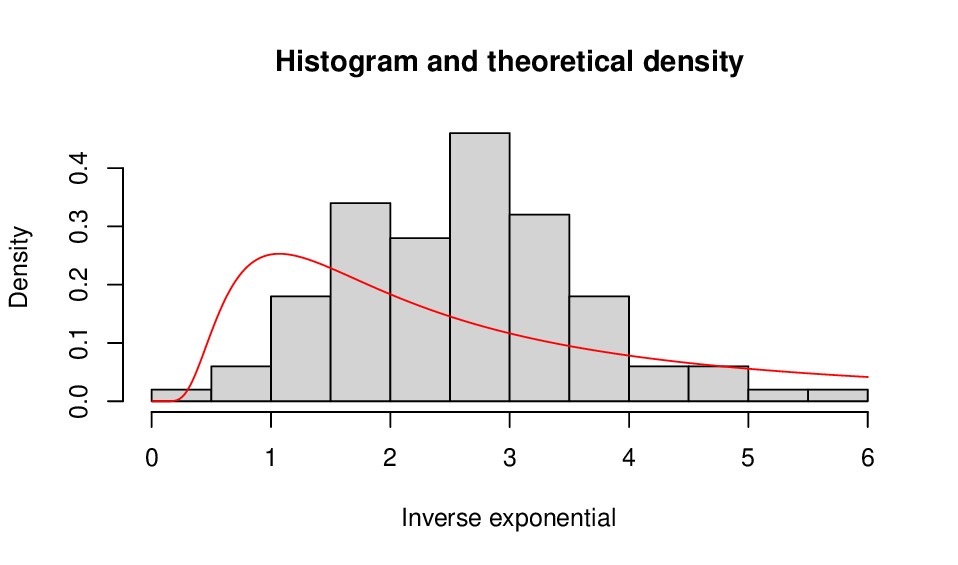}}
		\subfigure[]{\label{c1}\includegraphics[width=1.85in]{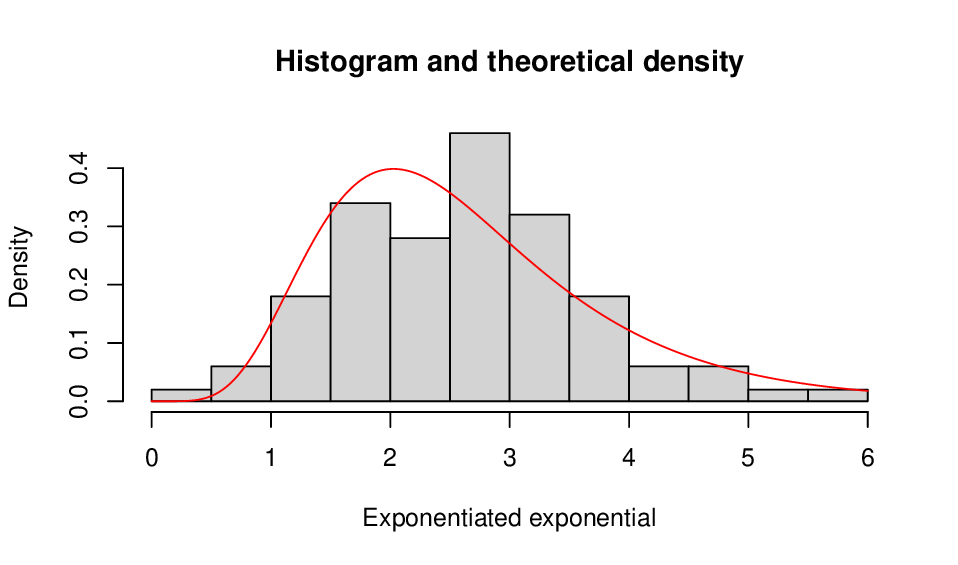}}
		\subfigure[]{\label{c1}\includegraphics[width=1.85in]{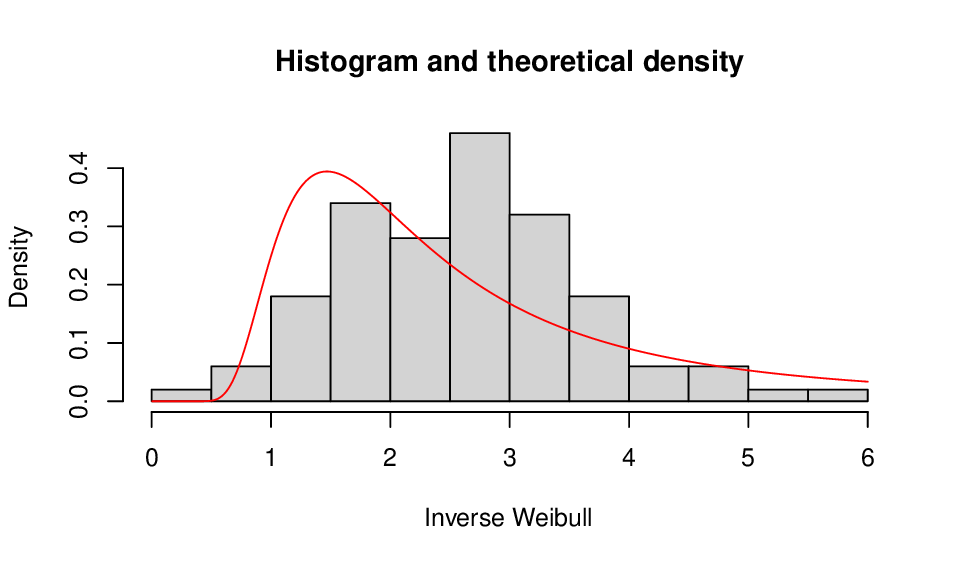}}
		\subfigure[]{\label{c1}\includegraphics[width=1.85in]{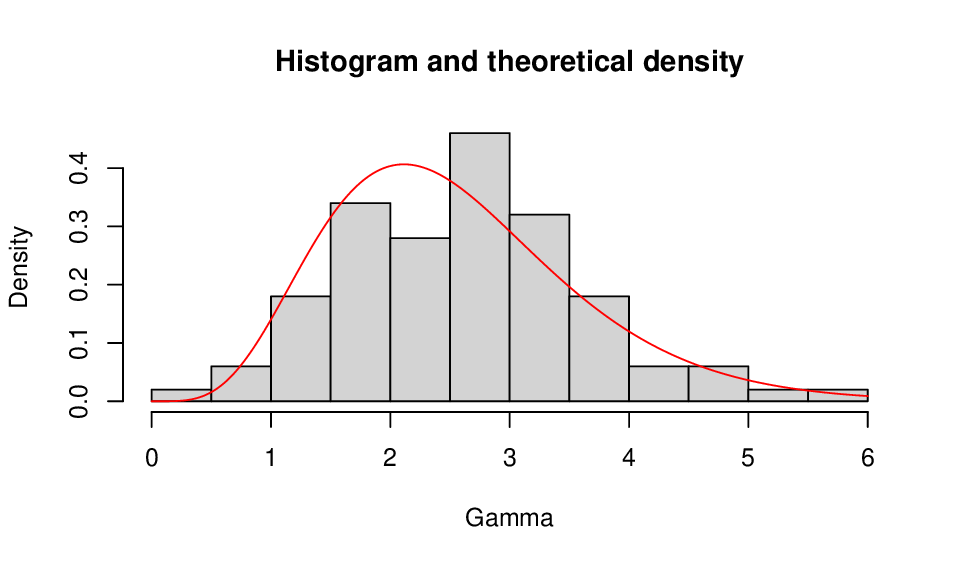}}
		\caption{The histogram and density plots for various distributions fitted to the assumed real data set.}
	\end{center}
\end{figure}

	\section{Conclusion}
	In this article, statistical inference for the LE
	distribution under PT-IHCS has been discussed.
	The MLEs are obtained. The forms of the MLEs can not be expressed explicitly. Thus, Newton-Raphson iterative method is employed to compute the MLEs of the parameters. In Bayesian study, independent and bivariate priors are considered. Three loss functions are utilized. The Bayes estimates are difficult to get in closed forms. They are of the form of the ratio of two integrals. To evaluate approximate Bayes estimates, Lindley's approximation technique is applied. The HPD credible intervals are obtained by using importance sampling method. In addition to the HPD credible intervals, two methods are employed to computed the asymptotic confidence intervals of the model parameters. Monte Carlo simulation study is performed to compare the performance of the proposed estimates. From the numerical study, it is observed that the Bayes estimates with respect to the independent prior distributions perform better than the other estimates in terms of the average values and the MSEs. Further, a real dataset is considered for illustrative purposes.\\
	\\
	\\
	\textbf{Acknowledgement:}
	 The authors would like to thank the Editor in Chief, an Associate Editor and anonymous reviewers for their positive remarks and useful comments.  The author S. Dutta, thanks the Council of Scientific and Industrial Research (C.S.I.R.
	Grant No. 09/983(0038)/2019-EMR-I), India, for the financial assistantship received to carry out this
	research work. Both the authors thanks the research facilities received from the Department of Mathematics, National Institute of Technology Rourkela, India.
	\\
	\\
		\textbf{ Disclosure Statement:} Both the authors declare that they do not have any conflict of interest. 
	
	\bibliography{myref}
	
	\appendix
	\section{Appendix}\setcounter{equation}{0}
	In Subsection $3.1$, Bayes estimates based on univariate prior are obtained and the posterior density function has a term $k_{1}$ which is expressed as
	\begin{eqnarray}\label{A1}
		 \nonumber k_{1}&=& \int_{0}^{\infty}\int_{0}^{\infty} \alpha^{D+a-1}
		\lambda^{D+c-1} e^{-b\alpha} e^{-\lambda(d-\sum_{i=1}^{D}x_{i})}
		\prod_{i=1}^{D} (g(x_i;\lambda))^{\alpha-1}
		[1+(g(x_i;\lambda))^{\alpha}]^{-(R_{i}+2)}\\
		 &~&\times [1+(g(T;\lambda))^{\alpha}]^{-R^{*}_{D}} d\alpha d\lambda.
	\end{eqnarray}
    Similarly, to obtain Bayes estimates based on bivariate prior  the posterior density function has a term $k_{2}$ which is expressed as
    \begin{eqnarray}\label{A2}
    	\nonumber k_{2}&= &\int_{0}^{\infty}\int_{0}^{\infty}  \alpha^{D} \lambda^{D+c-2} e^{-\lambda(d-\sum_{i=1}^{D}x_{i})}\prod_{i=1}^{D} (g(x_i;\lambda))^{\alpha-1}  [1+(g(x_i;\lambda))^{\alpha}]^{-(R_{i}+2)} \\
    	 &~& \times [1+(g(T;\lambda))^{\alpha}]^{-R^{*}_{D}} d\alpha~d\lambda.
    \end{eqnarray}
In Subsection $4.2$, the posterior density has a weight function of $\alpha$ and $\lambda$ which is expressed as
	\begin{eqnarray}\label{A6}
	\nonumber h(\alpha,\lambda)&=& \left[b-\sum_{i=1}^{D}\log
	 (g(x_i;\lambda))\right]^{-(D+a)}\left(d-\sum_{i=1}^{D}x_{i}\right)^{-(D+c)}\left[1+(g(T;\lambda))^{\alpha}\right]^{-R^{*}_{D}}\\
	&~&\times \prod_{i=1}^{D}
	\frac{\left[1+(g(x_i;\lambda))^{\alpha}\right]^{-(R_{i}+2)}}{g(x_i;\lambda)}.
	\end{eqnarray}.

\section{Appendix}\setcounter{equation}{0}
	In this part of the paper, the Lindley's approximation method is
	described. However, one may refer to \cite{Lindley} for further
	elaboration. Consider an arbitrary function of $\alpha$ and
	$\lambda$, say $\phi(\alpha,\lambda)$. The forms of the Bayes
	estimators (see Eqs. (\ref{eq3.4}), (\ref{eq3.5}) and (\ref{eq3.6}))
	involve expectations with respect to the posterior probability
	density function. The posterior mean of $\phi(\alpha,\lambda)$ is
	given by
	\begin{eqnarray}\label{a1}
		E(\phi(\alpha,\lambda)|\underline{x})=
		 \frac{\int_{0}^{\infty}\int_{0}^{\infty}\phi(\alpha,\lambda)e^{l(\alpha,\lambda|\underline{x})+P(\alpha,\lambda)}
			d\alpha
			d\lambda}{\int_{0}^{\infty}\int_{0}^{\infty}e^{l(\alpha,\lambda|\underline{x})+P(\alpha,\lambda)}
			d\alpha d\lambda},
	\end{eqnarray}
	where $l(\alpha,\lambda|\underline{x})$ is the log-likelihood
	function and $P(\alpha,\lambda)$ is the logarithm of joint prior
	distribution of $\alpha$ and $\lambda$. Using the method due to
	\cite{Lindley}, (\ref{a1}) can be approximated as
	\begin{eqnarray}
		E(\phi(\alpha,\lambda)|\underline{x})\approx \phi(\alpha,\lambda) + \frac{1}{2}\left(A+l_{30}B_{12}+l_{03}B_{21}+l_{21}C_{12}+l_{12}C_{21}+2P_{1}A_{12}+2P_{2}A_{21} \right),\nonumber\\
	\end{eqnarray}
	where $A=\sum_{i=1}^{2}\sum_{j=1}^{2} v_{ij}\tau_{ij}$ ,
	$\theta_1=\alpha$, $\theta_2=\lambda,$ $v_{i}=\frac{\partial
		\phi}{\partial \theta_{i}}$ , $v_{ij}=\frac{\partial^2
		\phi}{\partial \theta_{i}\partial \theta_{j}}$, $P_{i}=
	\frac{\partial P}{\partial \theta_{i}}$ , $A_{ij}=
	v_{i}\tau_{ii}+v_{j}\tau_{ji}$ ,
	$B_{ij}=(v_{i}\tau_{ii}+v_{j}\tau_{ij})\tau_{ii}$ , $C_{ij}=
	3v_{i}\tau_{ii}\tau_{ij}+v_{j}(\tau_{ii}\tau_{jj}+2\tau^{2}_{ij})$ ,
	$l_{ij}=\frac{\partial^{i+j}l}{\partial\theta^{i}_{1}\theta^{j}_{2}}$
	, $P=\log \pi_{3}(\theta_{1},\theta_{2})$ and $\tau_{ij}$ is the
	$(i,j)$th element of $\left[-\frac{\partial^2
		l}{\partial\theta^{i}_{1}\partial\theta^{j}_{2}}\right]^{-1}$, where
	$i,~j=1,~2$. For the present problem, the loglikelihood function can
	be written as
	\begin{eqnarray}
		l&=&D\log \alpha + D\log \lambda + \lambda \sum_{i=1}^{D} x_{i} + (\alpha -1)\sum_{i=1}^{D}\log (g(x_i;\lambda))- \sum_{i=1}^{D}(R_{i}+2) \log U\nonumber\\
		&~& - R^{*}_{D} \log V,
	\end{eqnarray}
	where $U\equiv U(\alpha,\lambda)=1+(g(x_i;\lambda))^{\alpha}$
	and $V\equiv V(\alpha,\lambda)=1+(g(T;\lambda))^{\alpha}$. Next,
	the required partial derivatives of $l$ with respect to $\alpha$ and
	$\lambda$ are presented.
	\begin{eqnarray*}
		\nonumber l_{10}&=&\frac{\partial l}{\partial \alpha}= \frac{D}{\alpha} + \sum_{i=1}^{D}\log (g(x_i;\lambda)) -\sum_{i=1}^{D}(R_{i}+2)\frac{U_{\alpha}}{U} -R^{*}_{D}\frac{V_{\alpha}}{V}\\
		\nonumber l_{01}&=&\frac{\partial l}{\partial \lambda}= \frac{D}{\lambda} + \sum_{i=1}^{D}x_{i} + (\alpha-1)\sum_{i=1}^{D}\frac{x_{i}e^{\lambda x_{i}}}{g(x_i;\lambda)} -\sum_{i=1}^{D}(R_{i}+2)\frac{U_{\lambda}}{U} -R^{*}_{D}\frac{V_{\lambda}}{V}\\
		\nonumber l_{20}&=&\frac{\partial^2 l}{\partial \alpha^2}= -\frac{D}{\alpha^2} -\sum_{i=1}^{D}(R_{i}+2)\left[\frac{UU_{\alpha\alpha}-U^{2}_{\alpha}}{U^2}\right] -R^{*}_{D} \left[\frac{VV_{\alpha\alpha}-V^{2}_{\alpha}}{V^2}\right]\\
		\nonumber l_{02}&=&\frac{\partial^2 l}{\partial \lambda^2}=
		-\frac{D}{\lambda^2}
		-(\alpha-1)\sum_{i=1}^{D}\frac{x^{2}_{i}e^{\lambda
				x_{i}}}{(g(x_i;\lambda))^2
		}-\sum_{i=1}^{D}(R_{i}+2)\left[\frac{UU_{\lambda\lambda}-U^{2}_{\lambda}}{U^2}\right]
		\\
		&~&-R^{*}_{D} \left[\frac{VV_{\lambda\lambda}-V^{2}_{\lambda}}{V^2}\right]\\
		\nonumber l_{11}&=&\frac{\partial^2 l}{\partial \alpha\partial \lambda}= \sum_{i=1}^{D}\frac{x_{i}e^{\lambda x_{i}}}{g(x_i;\lambda)} -\sum_{i=1}^{D}(R_{i}+2)\left[\frac{UU_{\alpha\lambda}-U_{\alpha}U_{\lambda}}{U^2}\right] -R^{*}_{D} \left[\frac{VV_{\alpha\lambda}-V_{\alpha}V_{\lambda}}{V^2}\right]\\
		\nonumber l_{30}&=&\frac{\partial^3 l}{\partial \alpha^3}=
		\frac{2D}{\alpha^3}-\sum_{i=1}^{D}(R_{i}+2)\left[\frac{U^2U_{\alpha\alpha\alpha}-
			3UU_{\alpha}U_{\alpha\alpha}+2U^{3}_{\alpha}}{U^3}\right]\\&~&
		-R^{*}_{D}\left[\frac{V^2V_{\alpha\alpha\alpha}
			-3VV_{\alpha}V_{\alpha\alpha}+2V^{3}_{\alpha}}{V^3}\right]\\
	\end{eqnarray*}
	\begin{eqnarray*}
		\nonumber l_{03}&=&\frac{\partial^3 l}{\partial \lambda^3}=(\alpha-1)\sum_{i=1}^{D}x^{3}_{i}\left[\frac{e^{\lambda x_{i}}(1+e^{\lambda x_{i}})}{(g(x_i;\lambda))^3}\right] -\sum_{i=1}^{D}(R_{i}+2)\left[\frac{U^2U_{\lambda\lambda\lambda}-3UU_{\lambda}U_{\lambda\lambda}+2U^{3}_{\lambda}}{U^3}\right]\\
		\nonumber &~& -R^{*}_{D}\left[\frac{V^2V_{\lambda\lambda\lambda}-3VV_{\lambda}V_{\lambda\lambda}+2V^{3}_{\lambda}}{V^3}\right]+\frac{2D}{\lambda^3}\\
		\nonumber l_{21}&=& \frac{\partial^3 l}{\partial\alpha^2\partial \lambda}= -\sum_{i=1}^{D}(R_{i}+2) \left[\frac{U(UU_{\alpha\alpha\lambda}-U_{\lambda}U_{\alpha\alpha}-2U_{\alpha}U_{\alpha\lambda})+2U^{2}_{\alpha}U_{\lambda}}{U^3}\right]\\
		\nonumber &~& -R^{*}_{D}\left[\frac{V(VV_{\alpha\alpha\lambda}-V_{\lambda}V_{\alpha\alpha}-2V_{\alpha}V_{\alpha\lambda})+2V^{2}_{\alpha}V_{\lambda}}{V^3}\right]\\
		\nonumber l_{12}&=& \frac{\partial^3 l}{\partial\alpha \partial \lambda^2}= -\sum_{i=1}^{D}(R_{i}+2) \left[\frac{U(UU_{\alpha\lambda\lambda}-U_{\alpha}U_{\lambda\lambda}-2U_{\lambda}U_{\alpha\lambda})+2U_{\alpha}U^{2}_{\lambda}}{U^3}\right]\\
		\nonumber
		&~&-R^{*}_{D}\left[\frac{V(VV_{\alpha\lambda\lambda}-V_{\alpha}V_{\lambda\lambda}-
			 2V_{\lambda}V_{\alpha\lambda})+2V_{\alpha}V^{2}_{\lambda}}{V^3}\right]-\sum_{i=1}^{D}x^{2}_{i}\left[\frac{e^{\lambda
				x_{i}}}{(g(x_i;\lambda))^2}\right]
	\end{eqnarray*}
	where
	\begin{eqnarray*}
		\nonumber U_{\alpha}&=&(e^{\lambda x_{i}}-1)^{\alpha} \log (g(x_i;\lambda))\\
		V_{\alpha}&=&(g(T;\lambda))^{\alpha} \log (g(T;\lambda))\\
		U_{\lambda}&=&\alpha x_{i}e^{\lambda x_{i}}(g(x_i;\lambda))^{\alpha-1}\\
		V_{\lambda}&=&\alpha Te^{\lambda T}(g(T;\lambda))^{\alpha-1}\\
		U_{\alpha\lambda}&=&x_{i} e^{\lambda x_{i}}(g(x_i;\lambda))^{\alpha-1}\left[\alpha\log (g(x_i;\lambda))+1\right]\\
		V_{\alpha\lambda}&=&T e^{\lambda T}(g(T;\lambda))^{\alpha-1}\left[\alpha\log (g(T;\lambda))+1\right]\\
		U_{\alpha\alpha}&=&(g(x_i;\lambda))^{\alpha}\left[\log (g(x_i;\lambda))\right]^2\\
		 V_{\alpha\alpha}&=&(g(T;\lambda))^{\alpha}\left[\log (g(T;\lambda))\right]^2\\
		U_{\lambda\lambda}&=&\alpha x^{2}_{i} e^{\lambda x_{i}}(g(x_i;\lambda))^{\alpha-2}(\alpha e^{\lambda x_{i}}-1)\\ V_{\lambda\lambda}&=&\alpha T^{2} e^{\lambda T}(g(T;\lambda))^{\alpha-2}(\alpha e^{\lambda T}-1)\\
		U_{\alpha\alpha\alpha}&=& (g(x_i;\lambda))^{\alpha}\left[\log (g(x_i;\lambda))\right]^3\\
		 V_{\alpha\alpha\alpha}&=& (g(T;\lambda))^{\alpha}\left[\log (g(T;\lambda))\right]^3\\
		U_{\lambda\lambda\lambda}&=& \alpha x^{3}_{i} e^{\lambda x_{i}} (g(x_i;\lambda))^{\alpha-3} \left[\alpha^2e^{2\lambda x_{i}}+(1-3\alpha)e^{\lambda x_{i}}+1\right]\\
		V_{\lambda\lambda\lambda}&=& \alpha T^{3} e^{\lambda T} (g(T;\lambda))^{\alpha-3} \left[\alpha^2e^{2\lambda T}+(1-3\alpha)e^{\lambda T}+1\right]\\
		U_{\alpha\alpha\lambda}&=& x_{i}e^{\lambda x_{i}}(g(x_i;\lambda))^{\alpha-1} \log (g(x_i;\lambda)) \left[\alpha\log (g(x_i;\lambda))+2\right]\\
		V_{\alpha\alpha\lambda}&=&Te^{\lambda T}(g(T;\lambda))^{\alpha-1} \log (g(T;\lambda)) \left[\alpha\log (g(T;\lambda))+2\right]\\
		U_{\alpha\lambda\lambda}&=&x^{2}_{i}e^{\lambda x_{i}}(g(x_i;\lambda))^{\alpha-2}\left[\alpha(\alpha e^{\lambda x_{i}}-1)\log (g(x_i;\lambda))+2\alpha e^{\lambda x_{i}}-1\right]\\
		V_{\alpha\lambda\lambda}&=&T^{2}e^{\lambda T}(g(T;\lambda))^{\alpha-2}\left[\alpha(\alpha e^{\lambda T}-1)\log (g(T;\lambda))+2\alpha e^{\lambda T}-1\right].
	\end{eqnarray*}

\end{document}